\documentclass[12pt,titlepage]{utarticle}

\usepackage{amsmath,amssymb,amsbsy,amstext}
\usepackage{graphicx}
\usepackage{amsfonts}
\usepackage{amssymb}
\usepackage{color}
\usepackage{float}

\numberwithin{equation}{section}
\newcommand{\be}{\begin{eqnarray}}
\newcommand{\bea}{\begin{eqnarray}}

\newcommand{\ee}{\end{eqnarray}}
\newcommand{\eea}{\end{eqnarray}}
\newcommand{\ba}{\begin{array}}
\newcommand{\ea}{\end{array}}

\newcommand{\Sos}{S_{\mathrm{on-shell}}}
\newcommand{\Sct}{S_{\mathrm{ct}}}

\newcommand{\ca}{A_{1,\mathrm{approx}}}

\newcommand{\au}{A_1}

\newcommand{\lbco}{La$_{2-x}$Ba$_x$CuO$_4$}
\newcommand{\lesco}{ La$_{1.8-x}$Eu$_{0.2}$Sr$_{x}$CuO$_{4}$}
\newcommand{\lnsco}{La$_{1.6-x}$Nd$_{0.4}$Sr$_{x}$CuO$_4$}

\newcommand{\A}{A}

\begin{document}

\baselineskip=15.5pt 

\title{A Striped Holographic Superconductor}

\author{Raphael Flauger
   \address{
      Department of Physics,\\
      Yale University,\\
      New Haven, CT 06520, USA\\
   }$^,$
   \address{
      Institute for the Physics and\\  Mathematics of the Universe\\
      The University of Tokyo\\
      Kashiwa, Chiba 277-8582, Japan
   }, Enrico Pajer $^3$ and Stefanos Papanikolaou
   \address{
      Department of Physics,\\ 
      Cornell University,\\
      Ithaca, NY 14853, USA\\
      {~}\\[.5cm]
      {\rm {\it e}-mail}\\[.1cm]
      \emailt{raphael.flauger@yale.edu}\\
      \emailt{ep295@cornell.edu}\\
      \emailt{sp682@cornell.edu}\\
   }
}
\preprint{IPMU10-0179}
\Abstract{We study inhomogeneous solutions of a 3+1-dimensional Einstein-Maxwell-scalar theory. Our results provide a holographic model of superconductivity in the presence of a charge density wave sourced by a modulated chemical potential. We find that below a critical temperature $T_c$ superconducting stripes develop. We show that they are thermodynamically favored over the normal state by computing the grand canonical potential. We investigate the dependence of $T_c$ on the modulation's wave vector, which characterizes the inhomogeneity. We find that it is qualitatively similar to that expected for a weakly coupled BCS theory, but we point out a quantitative difference. Finally, we use our solutions to compute the conductivity along the direction of the stripes.}

\maketitle


\section{Introduction}

Strongly coupled systems are often found at the heart of spectacular phenomena in condensed matter. Most prominently, they are expected to play a key role in the understanding of one of the most exciting scientific discoveries of the last thirty years, high temperature superconductivity. While some techniques have been devised over the years to model such systems, there is a scarcity of simple, tractable models on which to build our intuition. Recently, there has been a flurry of activity trying to fill in this gap using various developments of the gauge/gravity correspondence.

The main complexity that arises in strongly correlated electron systems, such as the high $T_c$ cuprates and the most recent iron pnictides, is the presence of ``competing orders"~\cite{vojta-review}: while the system may be a superconductor, it is not a homogeneous one and orders appear that are related to the breaking of the lattice symmetries, but are seemingly unrelated to superconductivity. Does the ubiquitous presence of such orders in strongly correlated superconductors have a deep connection to the very emergence of superconductivity and/or the magnitude of its critical temperature $T_c$? The answer to this question seems to be crucial for identifying the mechanism of superconductivity, at least in the cuprates. 
It is an important empirical observation, derived mainly from numerical calculations of Hubbard models~\cite{hubbard}, that inhomogeneity plays a much bigger role in these systems than in standard, weakly coupled superconductors. In fact, models have been proposed based on the coexistence of homogeneous superconductivity with charge density (CDW) or spin density waves (SDW), as well as models where the superconducting order parameter itself is modulated (pair density waves or PDW). Signatures of CDW have been reported in a variety of strongly correlated superconductors, most notably the hole-doped cuprates \lnsco, \lesco, and \lbco~\cite{vojta-review}. For these materials, the order can be consistently interpreted in terms of uni-directional SDW and CDW over a wide range of doping. Altogether, there is ubiquitous experimental evidence of ordered inhomogeneous structures in various places of the doping-temperature phase diagram of the cuprates, especially in the proximity of the superconducting phase. 

Regardless of whether strong coupling and inhomogeneity show us the path to the mystery of high $T_c$ superconductors or other interesting related systems, it is of great value to have simple computable models that incorporate these properties. In this paper we provide one model of this kind, in the hope that it might help improve our intuition about these key ingredients.

The AdS/CFT correspondence suggests the equivalence of certain gauge theories at conformal fixed points (CFT) with certain string theories on Anti-de-Sitter (AdS) backgrounds. There is a dictionary relating observables in the gauge theory to observables in the string theory. One of the key relations is between the curvature of the AdS background and the inverse of the gauge, (or rather 't Hooft) coupling of the gauge theory. At large 't Hooft coupling and large rank of the gauge group, the D-dimensional CFT becomes dual to gravity on D+1-dimensional AdS space, and the correspondence provides a tool to compute observables in a strongly coupled theory by solving classical equations of motion on a weakly curved background. Further developments of this idea allow one to include perturbations away from conformality and develop a richer structure on both sides of the duality. The use of this correspondence has allowed, in an unprecedented manner, the study and understanding of conformally invariant quantum systems in D dimensions. 

Over the past few years, a growing amount of effort has been devoted to applications of the AdS/CFT correspondence to strongly correlated condensed matter systems close to quantum critical points. A popular model that includes some important ingredients of a realistic system is the holographic superconductor~\cite{Hartnoll:2008kx,Hartnoll:2008vx}. It consists of a 3+1-dimensional Einstein-Maxwell-scalar theory in an AdS black hole background. At low temperatures, a phase transition to a superfluid state takes place \cite{Gubser:2008px}, in which the scalar field develops a non-vanishing expectation value, spontaneously breaking the gauge $U(1)$ symmetry. Physical properties such as transport coefficients can be studied and contrasted with other known systems like Bardeen-Cooper-Schrieffer (BCS) superconductors. 
The main conceptual ingredients of the holographic superconductor are strong coupling and proximity to an underlying conformal symmetry, which are believed to be crucial features of cuprates as well. On the other hand, inhomogeneity is not accounted for, although it certainly plays a role in the real systems.


In this paper, we would like to raise and attempt to answer a few questions: can one take holographic superconductors one step closer to real systems by considering inhomogeneous configurations? Can one account for charge or spin density wave (CDW or SDW) backgrounds? Are there any differences in the role of inhomogeneity played in the strongly coupled regime as opposed to the weakly coupled BCS case? How do physical properties such as $T_c$ and anisotropic conductivities compare with experimental data?

We address these questions by studying inhomogeneous solutions of the holographic superconductor with a modulated chemical potential of wave vector $Q$. Throughout this paper, we work in the probe approximation, neglecting the backreaction of the Maxwell and scalar field on the gravity background. The normal phase then consists of a CDW with the same modulation as the chemical potential and a vanishing order parameter. As the temperature is decreased the system undergoes a phase transition to a superconducting state with a non-vanishing \textit{modulated order parameter}. We find that various classes of solutions are possible. For the two simplest ones, the order parameter is real and the 3-dimensional gauge potential vanishes. The periodicity of the order parameter can then be taken to be the same as the one of the charge density, or it can be half of it. Here we concentrate on the former possibility, but we notice that the latter corresponds to a pair density wave (PDW) \cite{vojta-review}. We find that the CDW solutions are thermodynamically favored over the PDW solutions, but suspect that it may be possible to stabilize the PDW solutions in the presence of a magnetic field. We leave a careful study of these solutions for the future. Other solutions that correspond to less commonly studied configurations such as spiral pair density waves can also be found. The dependence of $T_c$ on the wave vector of the modulation (figure \ref{figTQ}) is qualitatively similar to the one expected for a weakly coupled BCS system \cite{martin05}. The two analyses have the same large and small $Q$ asymptotics for $T_{c}$ and both show the presence of a inflection point. The main difference is quantitative and concerns the functional dependence of $T_{c}$ on $Q$ in the $Q\rightarrow \infty$ limit.


We numerically investigate the superconducting state below $T_c$. We find that superconducting stripes develop and higher harmonics of the CDW are excited. We verify that this state has a lower grand canonical potential (the equivalent of the free energy, but in the grand canonical ensemble) than the normal CDW, as expected. Finally, we show results for the conductivity along the stripes, verifying the superconducting nature of the inhomogeneous phase. The conductivity in the orthogonal direction is substantially more complicated to compute and will be discussed in a future publication.  

The outline of this paper is as follows: in the next section, we introduce the gravitational system that we investigate, briefly describe its strongly coupled dual theory, and review the construction of a homogeneous holographic superconductor. In section~\ref{s:Tc}, we construct an inhomogeneous generalization corresponding to a CDW and show that it undergoes a phase transition at low temperatures. The resulting inhomogeneous superconducting phase is discussed in section~\ref{s:sc}. We present results for the grand canonical potential and the conductivity in sections \ref{s:fe} and \ref{s:cond}, respectively. A discussion of our results is given in section~\ref{s:c}. Finally, in Appendix~\ref{a:f}, we collect various useful forms of the equations of motion.


\section{The System: Einstein-Maxwell-Scalar Theory in 4D}\label{s:r}

The gravitational theory we consider is Einstein gravity in four dimensions with a negative cosmological constant $\Lambda=-3/L^2$, coupled to a $U(1)$ gauge field and a complex scalar field $\Psi$ which is charged under this $U(1)$ symmetry. Here $L$ is a length scale representing the AdS curvature radius. The action for this system is
\be\label{La}
S =\int d^4x\sqrt{-g}\left[ \frac1{16\pi G_N}\left(R + \frac{6}{L^2}\right) - \frac{1}{4} F^{ab} F_{ab} - g^{ab}(D_a\Psi)^*D_b\Psi - V(|\Psi|)\right]\,,
\ee
where $D_a=\partial_a-iqA_a$ and $a,b\in{t,x,y,z}$. The resulting equations of motion are given in \eqref{eq:eompsi}, \eqref{eq:eomA} and \eqref{eq:einst} of appendix \ref{a:f}. By the gauge/gravity duality \cite{Maldacena:1997re}, this system on an asymptotically anti-de Sitter spacetime is believed to be dual to some large N, strongly coupled, conformal gauge theory with a global $U(1)$ symmetry in 2+1 dimensions (for reviews see \cite{reviews}).\footnote{Since the $U(1)$ symmetry on the boundary is global, one is strictly speaking describing a superfluid as opposed to a superconductor, but one can imagine weakly gauging the symmetry, making a comparison with e.g.~BCS theory meaningful.}

For concreteness, we will work with a potential 
\be
V(|\Psi|)=-\frac2{L^2}|\Psi|^2\,,
\ee
corresponding to a scalar mass $m^2=-2/L^2$. This is the same mass as that of a conformally coupled scalar in AdS$_4$. Since this mass is above the Breitenlohner-Friedmann bound $m_{BF}^2 L^2=-9/4$, it does not imply an instability. In the context of the homogeneous holographic superconductor, other potentials have been studied (see e.g.~\cite{Horowitz:2008bn}), and it has been found that the results do not qualitatively depend on the choice of the potential (on the other hand see also \cite{Gubser:2009cg}). We expect this to be true for the striped holographic superconductors as well.

With this potential, the equations of motion~\eqref{eq:eompsi}-\eqref{eq:einst} possess three linearly independent scaling symmetries under which the various quantities transform as
\be
X\to\lambda^{\alpha_{i}} X
\ee
with the charges $\alpha_i$ given by
\begin{table}[h]
\begin{center}\label{t1}
\begin{tabular}{|c|c|c|c|c|c|c||c|}\hline
&$x^a$&$L$&$q$&$A_adx^a$&$\Psi$&$ds^2$&$G_N$\\\hline
$\alpha_1$&1&0&0&0&0   &0  &0   \\\hline
$\alpha_2$&0&1&-1&1&0  &2  &0\\\hline
$\alpha_3$&0&0&-1&1&1  &0 &-2\\\hline 
\end{tabular}
\end{center}
\end{table}

We can ignore the backreaction of the matter on the geometry if the invariant quantities such as $G_N|\Psi|^2$ are small. As usual, this can be achieved by sending $G_N$ to zero while keeping everything else fixed. Equivalently, using the third scaling symmetry, we can send the gauge potential and scalar field to zero while keeping $qA$ and $q\Psi$ fixed, corresponding to a limit in which the strength of the gauge interactions is infinitely larger than the gravitational interactions.
We will exclusively work in this limit in this paper and study the Maxwell and scalar fields in a fixed background. The superconductors we would like to describe are at finite temperature. So we take the background to be a planar Schwarzschild anti-de Sitter black hole which corresponds to studying the dual field theory at finite temperature. We choose coordinates so that the line element is

\begin{equation}\label{metricz}
ds^2=\frac{L^2}{z^2}\left[-h(z)dt^2+\frac{dz^2}{h(z)}+dx^2+dy^2 \right]\,\,\,\,\text{with}\,\,\,\,h(z)=1-\frac{z^3}{z_0^3}\,,
\end{equation}
where $z=z_0$ is the position of the horizon of the black hole. The Hawking temperature of this black hole is
\be
T=\frac{3}{4\pi z_0}\,.
\ee 

The equations of motion for the gauge and scalar field \eqref{eq:eompsi} and \eqref{eq:eomA} with the metric corresponding to the line element~\eqref{metricz} are of course still invariant under our scaling symmetries (provided we assign the same charges to $z_0$ as to the $x^a$). We choose to use them to set $z_0=1$, $L=1$, and $q=1$ to simplify the equations but always plot quantities that are invariant under these rescalings.

For later use, we give the equations of motion in these conventions written in components of the gauge field and the real and imaginary part of the scalar field in Appendix A.  


\subsection{Review of the Homogeneous Solutions}\label{s:hhsc}

A simple set of solutions to \eqref{eq:eompsi} and \eqref{eq:eomA} can be obtained \cite{Hartnoll:2008vx} by assuming that the fields do not depend on $x,y$ and $t$. In this case it is consistent to set $A_{z}=A_{x}=A_{y}=0$ and to take $\Psi\equiv z\psi/\sqrt{2}$ to be real, where the particular definition of $\psi$ has been chosen for later convenience. With this Ansatz, only two real fields are left: $\psi$ and $A_t$, which we denote by $\A$ for simplicity.

The equations of motion \eqref{eomA} and \eqref{eomphi} on the background \eqref{metricz} with the assumption of homogeneity and stationarity take the form\footnote{Once again, we used the symmetries summarized in table \ref{t1} to set $q=L=z_0=1$.}
\be \label{vic}
 h A_{zz}- \psi^2 A&=&0\,,\\
- h^2 \psi_{zz}+3z^2 h \psi_z+\left(hz- A^2\right) \psi&=&0\,.
\ee
A solution is determined once we specify four boundary conditions. For each equation, we fix the boundary condition at $z=1$, i.e.~the horizon of the black hole so that the solution is regular. This implies $A(1)=0$ and $\psi_z(1)=-\psi(1)/3$. We choose the third boundary condition by requiring that $\psi$ has a normalizable profile, which implies $\psi(0)=0$. The last boundary condition determines which ensemble we want to use. To see this, consider a solution of the equation of motion \eqref{Aeom} for $A$ and expand it near the AdS boundary as
\be
 A(z)=A^{(0)}+zA^{(1)}\,.
\ee
According to the AdS/CFT dictionary, a gauge symmetry in the bulk theory corresponds to a global symmetry in the boundary gauge theory. $A^{(1)}$ is mapped into the expectation value of the time-component of the current of the global symmetry, while $A^{(0)}$ corresponds to a source in the boundary Lagrangian. Since $A^{(1)}$ can be interpreted as a charge density, its canonical conjugate $A^{(0)}$ plays the role of a chemical potential. Fixing the charge density or the chemical potential corresponds to working in the canonical or grand canonical ensemble, respectively. We decide to fix the chemical potential $A^{(0)}$ and therefore work in the grand canonical ensemble. Notice that in any case, specifying a non-vanishing boundary condition for $A$ breaks explicitly the conformal invariance of the boundary theory (which is also broken by temperature).

We first look for solutions in which $\psi=0$, corresponding to a vanishing order parameter. These are dual to the normal, non-superconducting state of the boundary theory. The equation of motion for $\psi$ is trivially satisfied. We are left with only the simple linear differential equation
\be\label{Aeom}
A_{zz}=0\,.
\ee
If we choose a chemical potential $\mu$ by imposing the boundary conditions $A(0)=\mu$ we find the solution 
\be\label{bkg}
A(z)=\mu\left(1-z\right)\,.
\ee
For low chemical potential, or equivalently high temperature, only this normal solution with $\psi=0$ exists. Beyond a critical value $\mu_c$, a new solution appears \cite{Gubser:2008px} that has non-vanishing condensate, i.e.~$\psi\neq0$. It has a lower grand canonical potential than the normal solution. This solution is typically studied numerically, although some analytical approaches have been considered in \cite{Siopsis:2010uq,Gregory:2009fj,Herzog:2010vz,Basu:2008bh}. The study of the physical properties of the $\psi\neq0$ solution \cite{Hartnoll:2008kx,Hartnoll:2008vx} reveals that its dual state is some kind of superfluid. The results for the conductivity are given in figure \ref{f:cond} and will be thoroughly discussed in section \ref{s:cond}.


\section{Inhomogeneous Solutions: the Normal State and $T_c$}\label{s:Tc}

In this section, we begin our study of static inhomogeneous solutions of the system~\eqref{La}. For previous discussions of inhomogeneous holographic constructions see \cite{Nakamura:2009tf} and \cite{Aperis:2010cd}, which appeared as this work was being finalized. Since we are interested in (static) uni-directional charge density waves, we study solutions with non-trivial dependence in the $x$-direction that are independent of $y$ or $t$.
It is then consistent to look for solutions of \eqref{eomphi}-\eqref{eomAlast} with $A_z=A_y=A_x=\mathrm{Im}\Psi=0$, and the equations of motion with this Ansatz take the form\footnote{We again set $q=L=z_0=1$ using the symmetries of the system.}
\be
 h A_{zz}- \psi^2 A+A_{xx} &=&0\,,\label{eomAin}\\
- h^2 \psi_{zz}+3z^2 h \psi_z-h\psi_{xx}+\left(hz- A^2\right) \psi&=&0\,, \label{eomphiin}
\ee
where once again $\psi\equiv \sqrt{2}\mathrm{Re}\Psi/z$.

Similar to what was found in the homogeneous case, we find that for a monochromatic chemical potential a phase transition occurs at some critical temperature. In the gauge theory, we interpret this as a phase transition between a normal and a superfluid/superconducting state in the presence of a charge density wave (CDW). In this section, we study the system in the normal phase in which $\psi=0$, as well as just below $T_c$ where the condensate is small. This allows us to work at linear order in $\psi$. Since $\psi$ enters the equation for $A$ quadratically, the equation for the gauge field $A$ in this approximation becomes independent of the scalar, simplifying the problem. This allows us to present analytical results for $Q=0$ in subsection \ref{ss:an}. Beyond recovering the homogeneous system, these analytic results provide us with some intuition for the effect of inhomogeneities. Since were unable to find analytic results for the most interesting range of $Q$, we provide numerical results for $T_c$ in subsection \ref{ss:Tc}. 

Working at linear order in the scalar field allows us to find the critical temperature $T_c$ at which the normal state develops an instability towards the superconducting state $\psi\neq0$, but is of course insufficient to study the system away from this critical temperature. We return to the study of the superconducting solution away from $T_c$ in the next section. 


\subsection{The CDW Background}

A simple solution of \eqref{eomAin} and \eqref{eomphiin} is  found if $\psi=0$ and corresponds to the normal non-superconducting state. Equation \eqref{eomphiin} is trivially satisfied. For \eqref{eomAin} we have to specify two boundary conditions. As before, one boundary condition is forced upon us, $A(1,x)=0$, if we require regularity at the horizon of the black hole. The second boundary condition is the physically relevant one since it amounts to specifying the chemical potential (or the charge density if we decided to work in the canonical ensemble) of the boundary theory. We decide to consider a constant plus a sinusoidal (monochromatic) term imposing 
\be \label{bc}
\A(z=0,x)=\mu \left[(1-\delta)+ \delta \cos(Q x)\right]\,,
\ee
where $Q$ is the $x$-frequency of the sinusoidal term and $\delta$ controls the ratio between inhomogeneous and homogeneous amplitude of the pinning field $\A(0,x)$. $\delta=0$ corresponds to the homogeneous holographic superconductor reviewed in section \ref{s:hhsc}. Since \eqref{eomAin} is linear, we can superimpose solutions and look for a solution of the form
\be
\A(z,x)=\A_0(z)+\A_1(z)\cos(Qx),.
\ee
The homogeneous contribution to $\A$ is then simply given by
\be
\A_0(z)=\mu(1-\delta)(1-z)\,,
\ee
while the  $z$-dependence of  $\au(z)$ is governed by
\be 
\left(1-z^3\right)\au''&=&Q^2 \au\,,\label{simple}
\ee
and the boundary conditions are
\be
\au(1,x)=0\qquad\text{and}\qquad\au(0,x)=\mu\delta\cos(Qx)\,.
\ee
\begin{figure}
\centering
\includegraphics{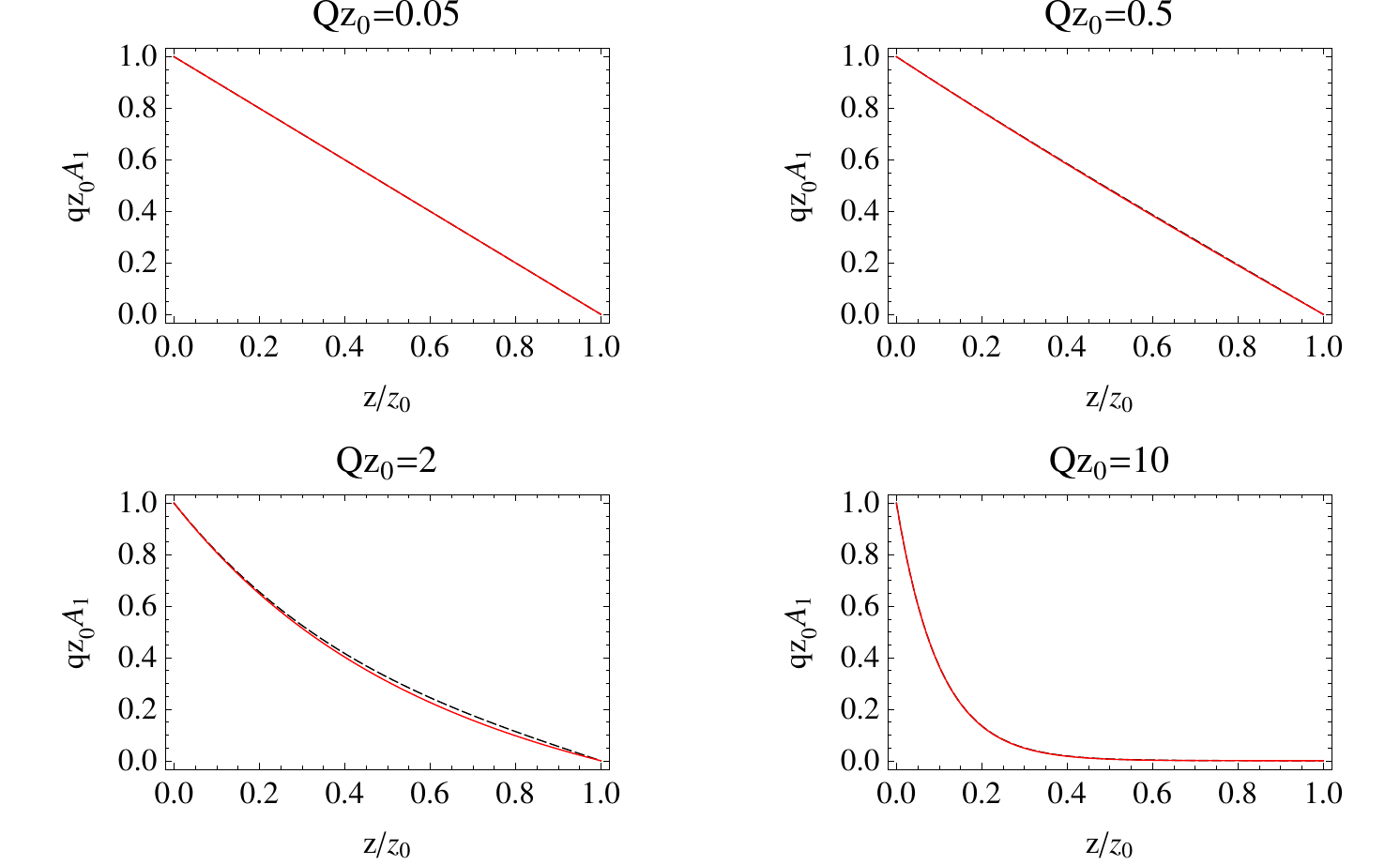}
\caption{The plot shows the numerical exact (continuous red line) and analytically approximate (dashed black line) solutions of \eqref{simple} for various values of $Q$. \label{f1}}
\end{figure}

Although we were not able to find an exact solution of \eqref{simple}, we notice that the numerical solution is very well approximated by the solution of the simpler equation $\au''=Q^2\au$, i.e.~
\be \label{app}
\au(z)\simeq \ca(z)\equiv\mu \delta \left[\cosh(Q z)-\coth(Q ) \sinh(Q z)\right]\,.
\ee
The behavior of the exact numerical solution of \eqref{simple} and the approximate solution \eqref{app} for various values of $Q$ is given in figure \ref{f1}. The agreement is excellent both for $Q\ll1$ and $Q\gg1$, and still reasonably good for $Q\simeq\mathcal{O}(1)$. To summarize, the inhomogeneous electric background in the normal state ($\psi=0$) is given by
\be \label{bkgA}
A_{\mathrm{normal}}(z,x)&\equiv&A_0(z)+\cos(Qx)\au (z)\nonumber\\
&=&\mu (1-\delta)(1-z)+ \cos(Qx) \au(z) \,,
\ee
where $\au(z)$ is the exact solution of \eqref{simple}.


\subsection{The Instability and the Critical Temperature}

As in the homogeneous case~\cite{Gubser:2008px}, the normal state with $\psi=0$ becomes unstable for large values of $\mu$ . To find the value of the chemical potential $\mu_c$ at which the instability develops, we look for marginally stable solutions of \eqref{eomphiin} in the background 
of \eqref{bkgA}. Once again, this is consistent because we are working at linear order in the scalar field $\psi$ in the present section so that the effect of the condensate on the gauge field is negligible. For the scalar field equation of motion, we find it convenient to work in Fourier space and look for a solution of the form
\be
\psi(x,z)=\sum\limits_{n=0}^\infty \psi_n(z)\cos(n Q x)\,.
\ee
The partial differential equation for the scalar field then turns into a system of coupled, linear, second order ordinary differential equations for the Fourier coefficients $\psi_n(z)$.  
The $\psi_{n}$ satisfy the system \eqref{hie}, which can be conveniently written as
\be\label{sys}
-\Psi''+\left(zh(z)-A_0^2\right)\Psi-A_1^2\mathbb{A}_{11}\Psi-2A_0A_1\mathbb{A}_{01}+h(z)Q^2\mathbb{Q}\Psi=0\,,
\ee
with $\Psi\equiv(\psi_0,\psi_1,\psi_2,\dots)$, and the matrices $\mathbb{A}_{11}$, $\mathbb{A}_{01}$ and $\mathbb{Q}$ given by
\begin{equation}
\mathbb{A}_{11}=\left(
\begin{array}{cccccccc}
\frac12 &  0      &\frac14 &    0    & 0 & 0 & 0 & \cdots\\
0       & \frac34 &  0     & \frac14 & 0 & 0 & 0 & \cdots\\
\frac12 &    0   & \frac12&  0  &\frac14 & 0 & 0 & \cdots\\
0 & \frac14 &    0   & \frac12&  0  &\frac14 & 0 &  \cdots\\
0 & 0 & \frac14 &    0   & \frac12&  0  &\frac14 &  \cdots\\
\vdots&\vdots&\vdots&\vdots&\vdots&\vdots&\vdots&\ddots
\end{array}
\right)\,,\\\qquad
\mathbb{A}_{01}=\left(
\begin{array}{cccccccc}
0       & \frac12 & 0  &    0    & 0 &  \cdots\\
1       & 0 &  \frac12 & 0 & 0 &  \cdots\\
0 &    \frac12& 0&\frac12  &0 &  \cdots\\
0 & 0 & \frac12   &0& \frac12&    \cdots\\
\vdots&\vdots&\vdots&\vdots&\vdots&\ddots
\end{array}
\right)\,,
\end{equation}
and
\begin{equation}
\mathbb{Q}=\left(
\begin{array}{cccccc}
0 & 0 & 0 & 0 & 0 &   \cdots\\
0 & 1 & 0 & 0 & 0 & \cdots\\
0 & 0 & 4 & 0 & 0 &  \cdots\\
0 & 0 & 0 & 9 & 0 &   \cdots\\
0 & 0 & 0 & 0 & 16 &   \cdots\\
\vdots&\vdots&\vdots&\vdots&\vdots&\ddots
\end{array}
\right)\,.
\end{equation}
We will now solve the system \eqref{sys} analytically in the limit $Q\rightarrow 0$. Then, in subsection \ref{ss:Tc}, we will present the general numerical results and discuss the critical temperature $T_c$.


\subsubsection{Analytical Solution for $Q=0$}\label{ss:an}

In the limit $Q\to0$, the profile of the $A$ field becomes
\be \label{Q=0}
\lim_{Q\rightarrow 0}A(z,x)=A_0(z)+A_1(z)\cos(Qx)=\mu(1-z)\left[(1-\delta)+\delta \cos(Qx)\right]\,,
\ee
where we have used the approximate analytical solution \eqref{bkgA}, which becomes exact in the $Q\rightarrow 0$ limit, but have kept the $\cos(Qx)$ because it couples different Fourier modes of the scalar field. Using \eqref{Q=0}, the system of equations for the Fourier coefficients is
\be
-\Psi''+\left(\frac{z}{z_0^3}h(z)\right)\mathbb{I} \Psi-\mu^2(1-z)^2\left[(1-\delta)^2\mathbb{I}+\delta^2\mathbb{A}_{11}+2(\delta-\delta^2)\mathbb{A}_{01}\right]\Psi=0\,.
\ee
Notice that in this limit all the Fourier components have the same $z$-dependence, so that it becomes natural to look for a solution of the form
\be
\Psi(z)=\mathbf{v}\psi_\lambda(z)
\ee
where $\mathbf{v}$ is an eigenvector of the matrix $(1-\delta)^2\mathbb{I}+\delta^2\mathbb{A}_{11}+2(\delta-\delta^2)\mathbb{A}_{01}$ satisfying 
\be
\left[((1-\delta)^2\mathbb{I}+\delta^2\mathbb{A}_{11}+2(\delta-\delta^2)\mathbb{A}_{01}\right]\mathbf{v}=\lambda\mathbf{v}\,.
\ee
The function $\psi_\lambda$ then satisfies
\be
-\psi_\lambda''+\left(\frac{z}{z_0^3}h(z)\right)\psi_\lambda-\mu^2(1-z)^2\lambda\psi_\lambda=0\,.
\ee
We see that the solution with the largest eigenvalue is the first to become tachyonic. To find this eigenvalue, notice that the matrix $(1-\delta)^2\mathbb{I}+\delta^2\mathbb{A}_{11}+2(\delta-\delta^2)\mathbb{A}_{01}$ is column-stochastic, i.e. the entries in each of its columns add up to one. This implies that its largest eigenvalue is $\lambda=1$. The corresponding eigenvector is 
\begin{equation}\label{vici}
v=\left(
\begin{array}{c}
\frac12\\
1\\
1\\
1\\
1\\
\vdots
\end{array}
\right)\,.
\end{equation}
These are the Fourier coefficients of $\pi\delta(x)$, i.e. for any periodic function
\be 
f(x)&=&\frac{a_0}{2}+\sum_{n=1}^{\infty} a_n \cos(nx) + b_n \sin(nx)\,,\\
a_n&\equiv& \frac{1}{\pi}\int_{0}^{2\pi} f(x) \cos (nx) dx\,, \quad n\geq 0\,,\\
b_n&\equiv& \frac{1}{\pi}\int_{0}^{2\pi} f(x) \sin (nx) dx\,, \quad n\geq 1\,,\\
\ee
one finds
\be
\int_0^{2\pi}dx\,f(x)\delta(x)=f(0)\,,
\ee
for
\be 
\pi \delta(x)=\frac12+\sum_{n=1}^{\infty}\cos(n x)\,.
\ee
The function $\psi_{\lambda=1}$ corresponding to \eqref{vici} satisfies
\be
-\psi_{\lambda=1}''+\left[\frac{z}{z_0^3}h(z)-(1-z)^2\mu^2\right]\psi_{\lambda=1}=0\,.
\ee
This is the same equation one finds in the homogeneous case. So we see that in the limit $Q\rightarrow 0$, independent of $\delta$, we find the same critical temperature as in the homogeneous system. We will call this temperature $T_{c,h}$. 
This property can be understood intuitively as follows. For small $Q$ the $x$ derivatives in \eqref{eomphiin} are irrelevant, so the system is effectively equivalent to a collection of homogeneous systems each one at a fixed $x$, with chemical potentials ranging from $\mu(1-2\delta)$ for $x=\pi/Q$, up to $\mu$ for $x=0$. Therefore, for any $\delta$ there is an (isolated) homogeneous system with chemical potential $\mu$ , sitting at $x=0$, which will go through a phase transition precisely at $(T/q\mu)_{c,h}$.
In addition to this result for $T_c$, which was expected, we can learn a few more things from this analytical computation. For non-zero $Q$, the higher Fourier modes are very costly and will be suppressed. One thus sees that the effect of the coupling to the modulated gauge potential tends to excite all the Fourier modes democratically while the gradients tend to suppress the high-$n$ modes. This suggests that it will be a good approximation to set to zero the Fourier modes above some $n_{max}$ which will depend on $Q$ and the desired degree of accuracy. 


\subsubsection{Numerical Results for $T_c$}\label{ss:Tc}

\begin{figure}
\includegraphics[width=\textwidth]{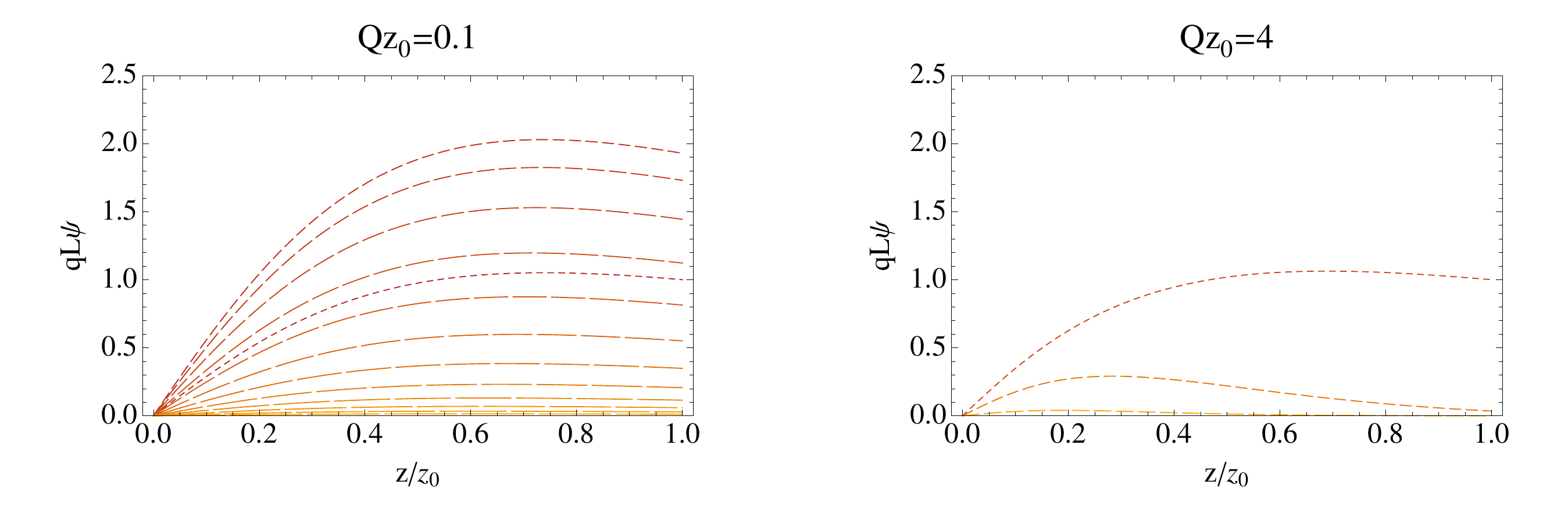}
\caption{We show the $z$ profiles of $\psi_n$ for the first few $n$ and two different values of $Q$ at the phase transition ($T/(q\mu)=0.0269$ for $Q=4$ and $T/(q\mu)=0.0575$ for $Q=1/10$). Dashed lines with longer dashes correspond to larger $n$. It is evident that $\psi_n$ are suppressed for larger $n$ and more so for larger $Q$. This results are obtained solving the (truncated) linear system \eqref{sys}, therefore the normalization of $\psi_n$ is arbitrary. \label{f2}}
\end{figure}
\begin{figure}
\includegraphics[width=\textwidth]{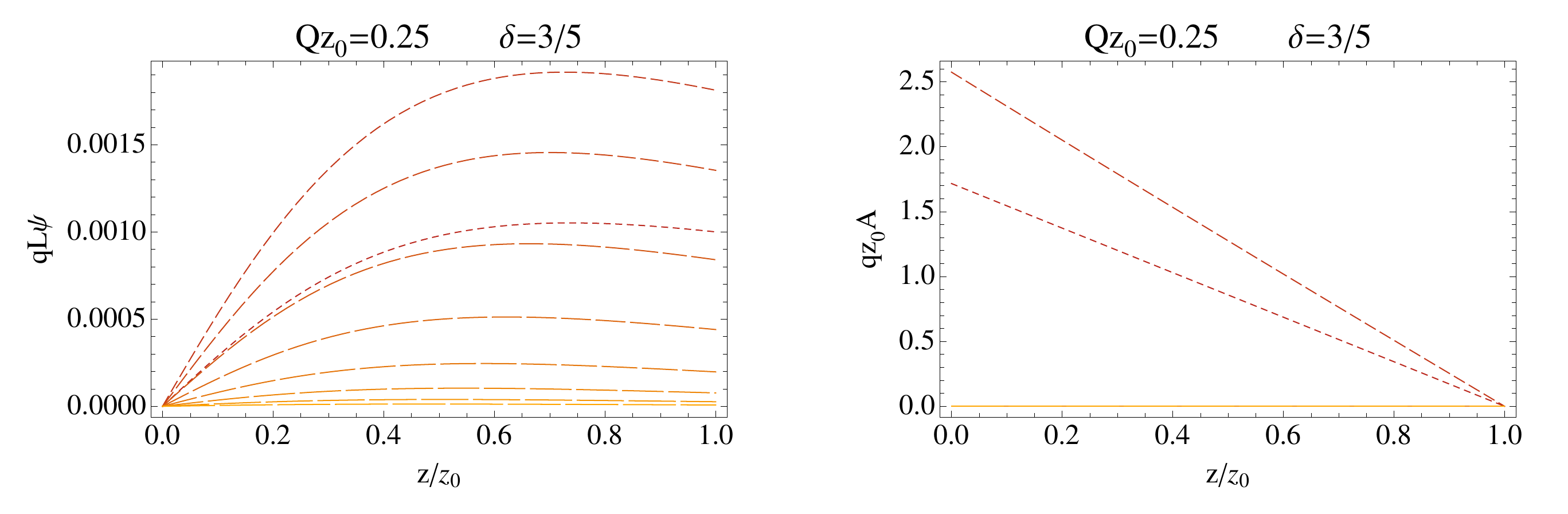}
\caption{The plots show the $z$ profiles of $\psi_n$ and $A_n$ obtained from the (truncated) full system \eqref{sys2}, for the first few $n$ and $\delta=3/5$, $Q=1/4$ and $T/(q\mu)=0.0556$ which is very close to the critical temperature. Again, dashed lines with longer dashes correspond to larger $n$. The $A_n$ profiles show that close to $T_c$ only the zeroth and first harmonics of $A$ are excited, which confirms that the backreaction of $\psi$ is negligible. \label{f2b}}
\end{figure}
In the following we present the numerical solutions of the system \eqref{sys} and discuss the critical temperature. Let us start by describing the numerical method we use. As we noticed, the effect of $\mathbb{Q}$ in \eqref{sys} is to suppress $\psi_n$ for large $n$. For a given momentum $Q$ and precision $\epsilon$, there is an $n_{max}(Q,\epsilon)$ such that all $\psi_{n}$ for $n>n_{max}(Q,\epsilon)$ contribute to $\psi$ less than $\epsilon$ and can be neglected. This is evident in figure \ref{f2} where the first $\psi_n$ are plotted for two different values of $Q$. This means we are allowed to truncate the hierarchy \eqref{sys}, and we are left with a finite number of $n_{max}(Q,\epsilon)$ coupled linear ordinary differential equations, which we solve numerically in Mathematica. In all the numerical results we present we have fixed the precision to be $10^{-3}$. Then, for example taking $Q=1/10$, we need to keep as many as $14$ modes, while for $Q=4$ only the first $3$ are necessary to achieve the required precision.

\begin{figure}
\centering
\includegraphics[width=6in]{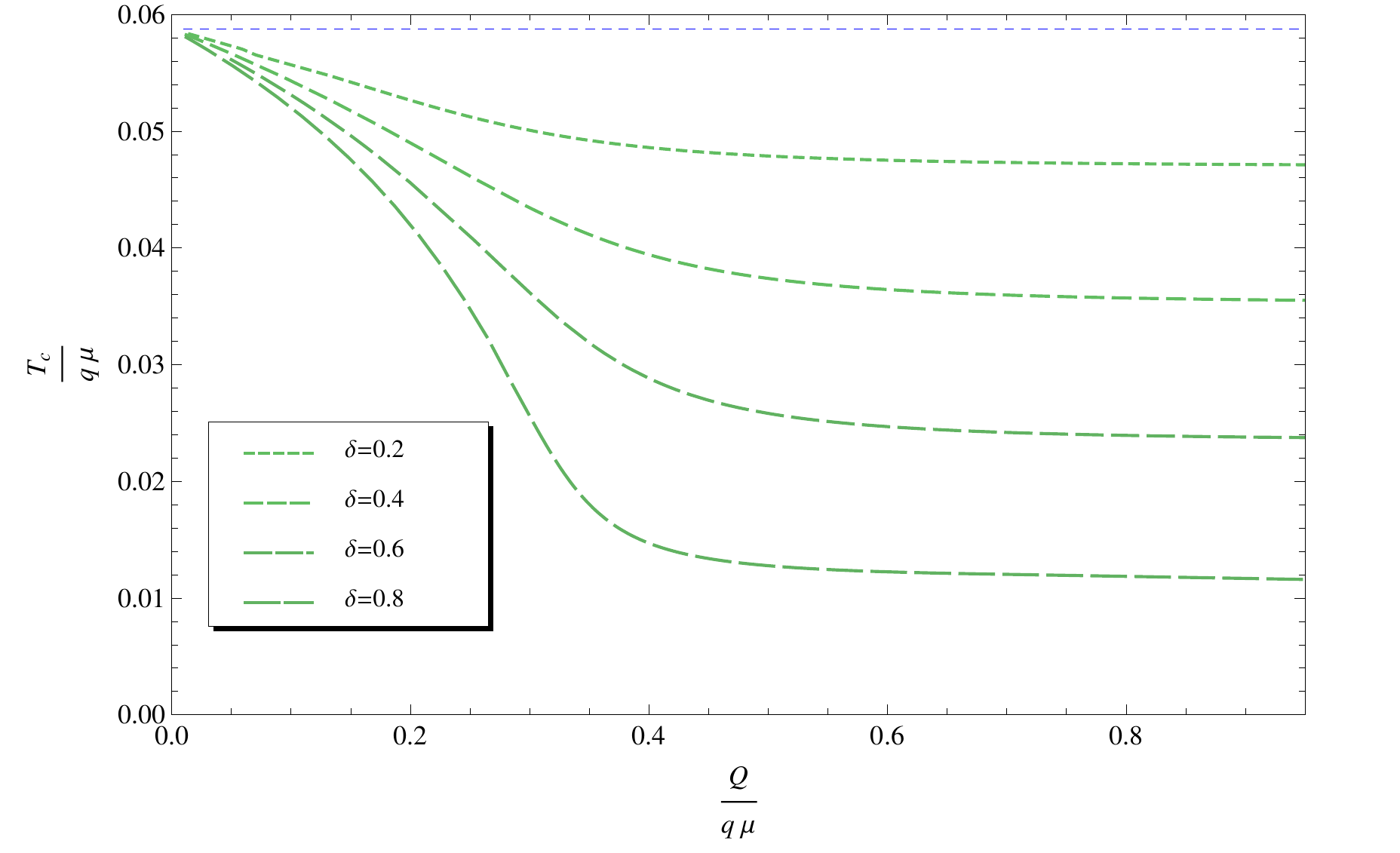}
\caption{The plot shows the critical temperature as function of Q for $\delta=0.2,0.4,0.6,0.8$.\label{figTQ}}
\end{figure}

Once we have solutions of \eqref{sys}, the most interesting physical property we can compute is the critical temperature $T_c$ at which the normal state becomes unstable. Let us pause to explain how we extract this quantity from the solutions of \eqref{sys}. Let us choose some fixed $0\leq\delta\leq 1$ and $Q\geq0$. Then our theory has four parameters $( L,q,z_0,\mu)$. Using the three symmetries in table \ref{t1}, we can set three of them to an arbitrary value and we once again set $L=z_0=q=1$.
Then we numerically solve \eqref{sys} by integrating from the horizon at $z=1$ out to the boundary at $z=0$. To set the initial conditions for this integration, we solve the equations as a power series near the horizon. If we impose regularity at the horizon, we will have one free coefficient for $A_0$, one for $A_1$, and one undetermined coefficient for each of the $n_{max}+1$ Fourier modes of $\psi$ giving a total of $n_{max}+3$ coefficients. Since the equation for $\psi$ is linear, the overall normalization is unphysical and we can set the coefficient for, say, $\psi_0$ to unity. The remaining $n_{max}+2$ coefficients are chosen so that $\psi_n(0)=0$ for all $n_{max}+1$ modes and $\delta A_0(0)=(1-\delta)A_1(0)$. We then have $n_{max}+2$ equations for the $n_{max}+2$ coefficients, which fixes all of them. In practice, we do not integrate to $z=0$ but to $z=10^{-8}$ and match the solution there to a power series of the solution of our~\eqref{sys} near the boundary to extract $A_0(0)$, $A_1(0)$ and $\psi_n(0)$. The resulting $\mu=A_0(0)/(1-\delta)$ is then $\mu_c$. The solution is unique if we require our Fourier coefficients $\psi_n(z)$ to have no zeros between the horizon and the boundary. The solutions with one or more zeros have higher $\mu$ and thus lower $T$ and are not relevant for us.
This $\mu$ is of course not invariant under the symmetries in table \ref{t1}, but the combination $T/(q\mu)$ is, so we can compute it for an arbitrary choice of those rescalings. This quantity is useful since variations of it can be interpreted, upon an appropriate rescaling, as variations of $T$ at constant $\mu$ and $q$. This is the quantity we will use in all of our plots involving temperature. 

In figure \ref{figTQ} we have plotted the critical temperature, obtained as explained above, as a function of $Q$ for a few different values of $\delta$. We notice the following features:
\begin{itemize}
\item The limit of $(T/q\mu)_c $ for $Q\rightarrow0$ is independent of $\delta$ and it corresponds to the critical value of a homogeneous superconductor with chemical potential $\mu$, which we denote by $(T/q\mu)_{c,h}$. This nicely agrees with our computation and discussion in the last subsection. 
\item The limit of $(T/q\mu)_c $ for $Q\rightarrow \infty$ depends on $\delta$ simply as 
\be 
\lim_{Q\rightarrow \infty}\left(\frac{T}{q\mu}\right)=(1-\delta)\left(\frac{T}{q\mu}\right)_{c}\Big|_{Q=0}=(1-\delta)\left(\frac{T}{q\mu}\right)_{c,h}\,.
\ee
This can be explained as follows. For very large $Q$, $A_1(z)$ goes to zero for any $z\neq0$, as can be seen in figure \ref{f1}. Then the background experienced by $\psi$ is that of the $x$-homogeneous field $A_0=\mu(1-\delta)(1-z)$ almost everywhere. Furthermore, as explained before, for $Q\to\infty$ only $\psi_0$ is non-zero. Hence the system is equivalent to a homogeneous one with chemical potential $\mu(1-\delta)$ and therefore experiences a phase transition at $(1-\delta)(T/q\mu)_{c,h}$.
\item One might ask how $(T/q\mu)_c (Q,\delta)$ reaches its asymptotic value for $Q\rightarrow\infty$. Since $\au$ vanishes exponentially in $Qz_0$ in the $Q\rightarrow \infty$ limit for any $z\neq0$, we expect it to behave as 
\be \label{exp}
\lim_{Q\rightarrow \infty}\left(\frac{T}{q\mu}\right)\simeq (1-\delta)\left(\frac{T}{q\mu}\right)_{c,h}+ e^{-Qz_0}\delta \left(\frac{T}{q\mu}\right)_{c,h}\,,
\ee
plus terms of higher order in $e^{-Qz_0}$ which become negligible for large $Q$. We confirm that this behavior is consistent with that of our numerical results by performing various fits.
\end{itemize}

It is interesting to compare the functional dependence of $T_c$ on $Q$ for our striped holographic superconductor with that of a weakly coupled BCS superconductor studied in \cite{martin05} using a mean-field approximation. Since we choose a simple consistent ansatz in which the order parameter is real, our analysis does not account for phase fluctuations and phase coherence, so we should compare with the continuous line in figure 1 of \cite{martin05}. The two results look extremely similar. Both the $Q=0$ and $Q=\infty$ asymptotics are the same and in both cases there is an inflection point at finite $Q$. The main difference consists in how the $Q=\infty$ limit is reached. Expanding (4) of \cite{martin05} for large $Q$ and translating into our notation for convenience, we find
\be 
\mathrm{BCS:}\lim_{Q\rightarrow \infty}T_c\simeq (1-\delta)T_{c,h}+ \frac{\delta}{\log Q} T_{c,h}+\mathcal{O}\left(\frac{1}{(\log Q)^2}\right)\,,
\ee
to be contrasted with the exponential behavior of \eqref{exp}. 
It would be interesting to generalize our analysis to include phase fluctuations and compare the results with the expectations of \cite{martin05}.


\section{Inhomogeneous Solutions: the Superconducting State}\label{s:sc}

In this section we study the full system of coupled non-linear partial differential equations in \eqref{eomphiin} and \eqref{eomAin}. We start by expanding both $A$ and $\psi$ in Fourier modes as
\bea 
\psi(x,z)&=&\sum\limits_{n=0}^\infty \psi_n(z)\cos(n Q x)\,,\\
A(x,z)&=&\sum\limits_{n=0}^\infty A_n(z)\cos(n Q x)\,,
\eea 
\begin{figure}
\includegraphics[width=6in]{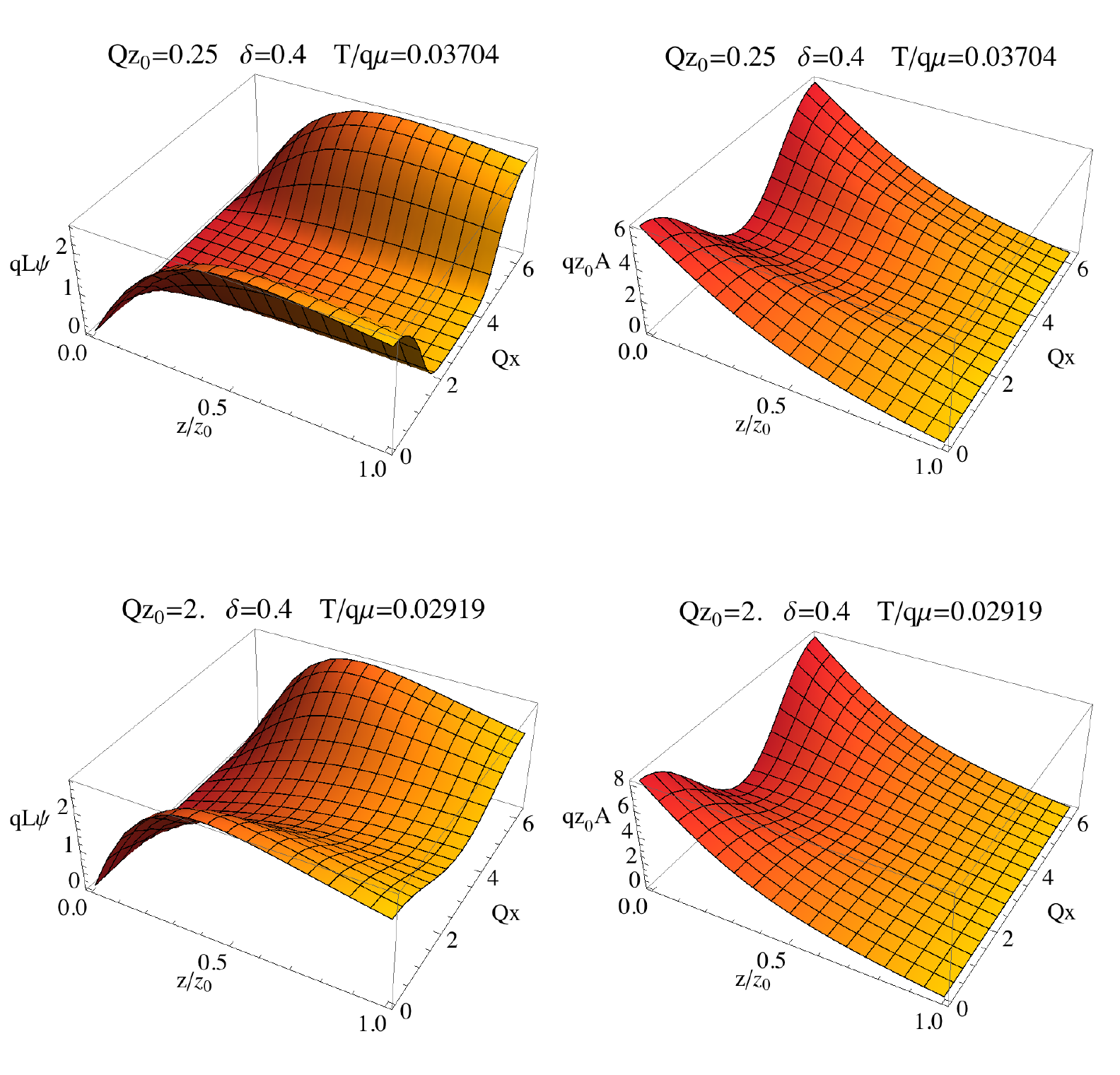}
\caption{The plots show the $z,x$ profile of $A$ and $\psi$ for $Q=.25$ and $Q=2$. The striped pattern of condensation of the bulk scalar field $\psi$ is evident in both cases, but more pronounced for lower $Q$'s.\label{figprofile}}
\end{figure}
\begin{figure}
\includegraphics[width=\textwidth]{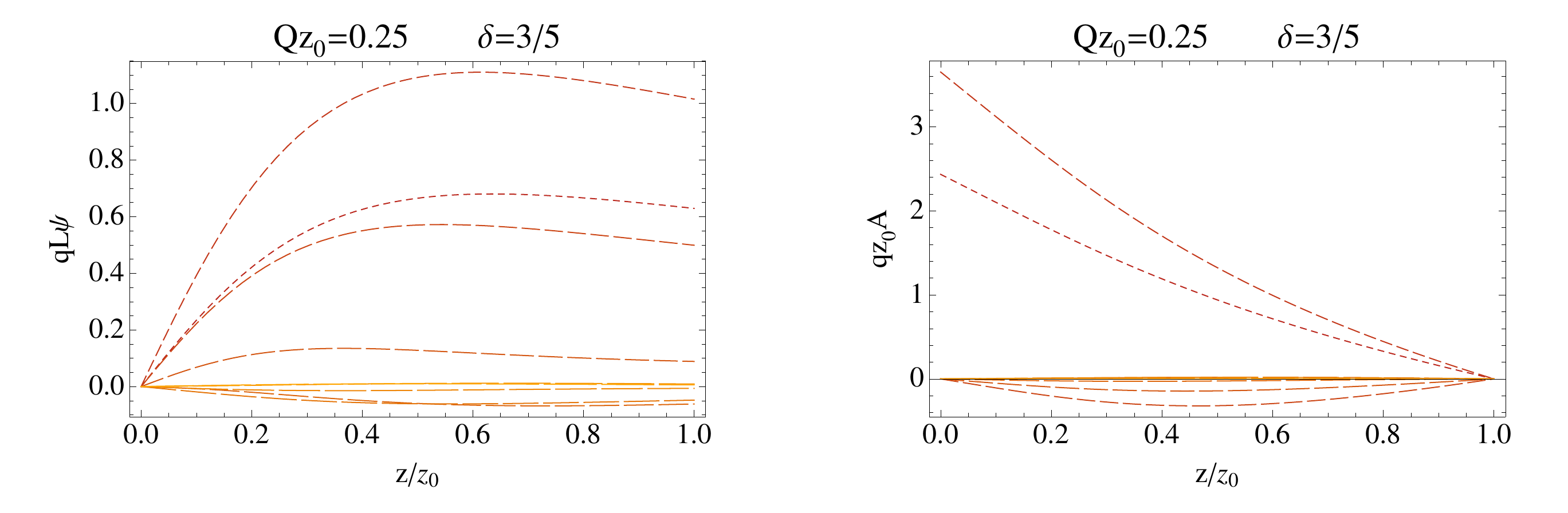}
\caption{The left and right plots show the first few $\psi_n$ and $A_n(z)$, respectively, obtained from the (truncated) full system \eqref{sys2} for $Q=1/4$, $T/(q\mu)=$ and $\delta=3/5$. Longer dashes correspond to larger $n$. The Fourier modes $A_n$ for $n>1$ are excited by the backreaction of the scalar condensate. \label{f:an}}
\end{figure} 
\begin{figure}
\includegraphics[width=\textwidth]{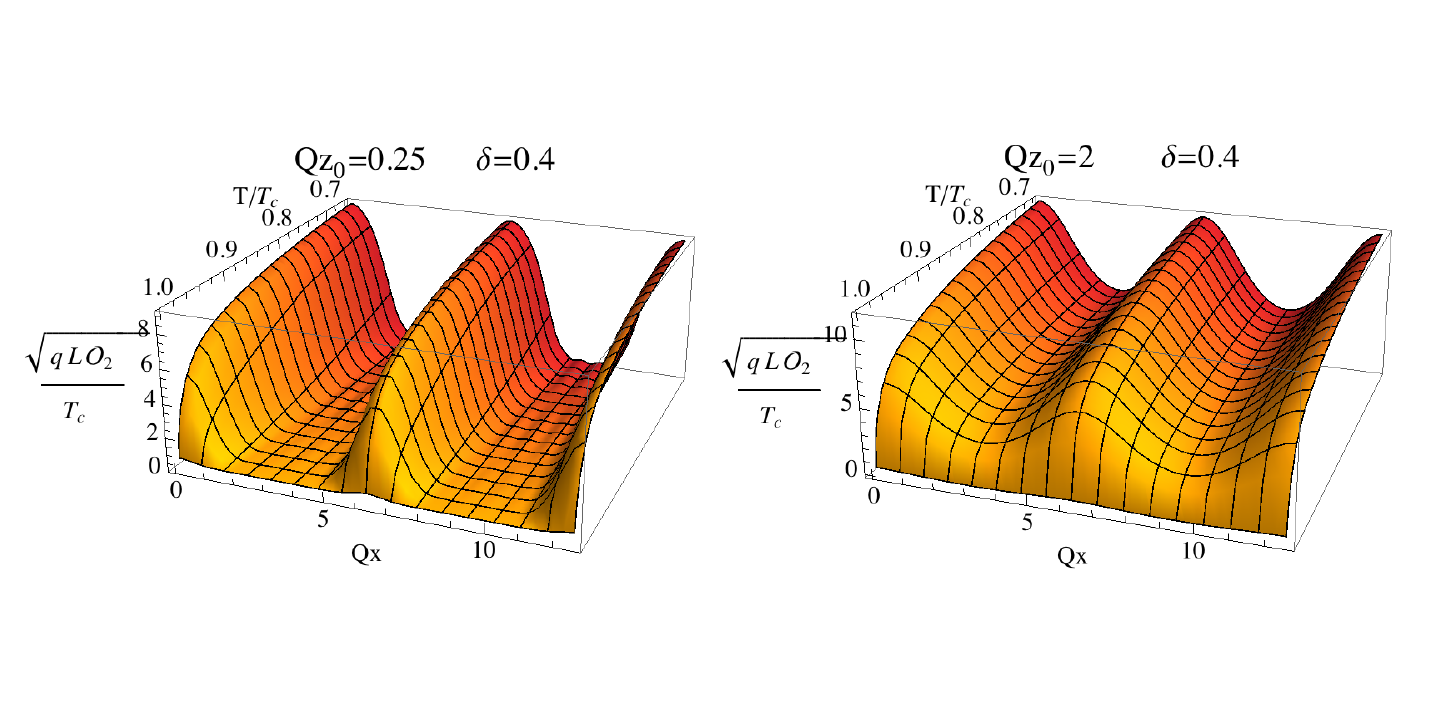}
\caption{The dependence of the superconducting order parameter $\mathcal{O}_2\equiv\partial_z\psi(0,x)$ on temperature and $x$ is plotted for $Q=1/4$ and $Q=2$ with $\delta=2/5$. We plot two periods in $x$ to better visualize the superconducting stripes, which have the same periodicity as the CDW. Notice that for small $Q$ the system is superconducting only in the stripes since the order parameter is very small in between. For larger values of $Q$, the system is stiffer in the $x$ direction and therefore the stripes are less pronounced. \label{f:stripes}}
\end{figure}
and plugging this back into \eqref{eomphiin} and \eqref{eomAin}. Then, by projecting onto each Fourier mode, we obtain a hierarchy of coupled non-linear ordinary differential equations for $A_n(z)$ and $\psi_n(z)$. As before, we can truncate the expansion at some $n_{max}(Q,\epsilon)$, which we again choose so that the modes $\psi_{n_{max}}$ and $A_{n_{max}}$ are a factor of $10^{-3}$ smaller than the $\psi_n$ or $A_n$ with the largest amplitude. The hierarchy is too long to be written out here, but for the convenience of the reader we include a few lines of Mathematica code \eqref{sys2} in the appendix that can compute it. To solve the hierarchy, we again integrate it from the horizon out to the boundary. The boundary conditions at the horizon are again obtained from a power series of the solution near the horizon. After setting $A_n(1)=0$ to ensure regularity, one has $2n_{max}+2$ undetermined coefficients.  We require $\psi_n(0)=0$ for the $(n_{max}+1)$ Fourier modes of the scalar field in order to have a normalizable profile. We also impose $A_0(0)=\mu(1-\delta)$, $A_1(0)=\mu\delta$ and $A_n(0)=0$ for $1<n\leq n_{max}$. This provides $2n_{max}+2$ equations for the $2n_{max}+2$ Fourier coefficients. Again, we only integrate to $z=10^{-8}$ and match to a power series expansion of the solution near the boundary to extract these quantities. As before, we choose the solution that has no zeros in the z-profile, which gives rise to a unique solution. The fact that the truncation is a good approximation can be seen in figure \ref{f:an} where we show a solution of the full system \eqref{sys2} for $A_n(z)$ and $\psi_n$ for the first few $n$. It is clear that both $A_n$ and $\psi_n$ become smaller and smaller as $n$ increases.


In figure \ref{figprofile} we show $\psi(z,x)$ and $A(z,x)$ as obtained from our numerical computation with the specified values of $T,\delta$ and $Q$. At some fixed subcritical $T<T_c$, as $Q$ is gradually increased (keeping everything else fixed), there is a smooth crossover between two qualitatively distinct types of profiles of $\psi$. For $Q< 1$, $\psi$ has a sharp striped condensation pattern in the bulk: the scalar field has a large expectation value in correspondence of the maxima of the CDW, while it is vanishingly small in between as can be seen in the top half of figure \ref{figprofile}. This behavior carries over to the order parameter of the boundary theory, i.e.~$\mathcal{O}_2(x)=\partial_z\psi(0,x)$, as we show on the left of figure \ref{f:stripes}. The boundary system has sharp superconducting stripes, well separated by stripes of normal phase. The situation changes continuously as we increase $Q$. For $Q>1$, the system is stiff enough in the $x$ direction that bulk condensation takes place everywhere at approximately the same temperature, although modulations are still clearly visible (bottom of figure \ref{figprofile}). This translates into the fact that the boundary system is  well inside the superconducting phase everywhere, but the order parameter is modulated as we show on the right of figure \ref{f:stripes}.


\section{The Grand Canonical Potential}\label{s:fe}

In this section we study the grand canonical potential $\Omega$ for the system. This is the Legendre transform of the free energy $F$, which is the thermodynamic potential in the canonical ensemble. Before proceeding, let us be pedantic and spell out this difference in detail. The natural variables for $F$ are $(T,\rho)$, while those for $\Omega$ are $(T,\mu)$, where $\mu$ is the chemical potential for $\rho$. Using the AdS/CFT dictionary, in our computation $A^{(0)}$ plays the role of the chemical potential $\mu$, since it perturbs the Lagrangian of the boundary theory, while $A^{(1)}$ plays the role of the (charge) density $\rho$, since it is the expectation value of the time component of a current, i.e. a density. We have decided to give our results in terms of $T/(q\mu)$ which is invariant under all the symmetries in table \ref{t1}. All our plots are directly comparable with any other convention for fixing three out of the four parameters $\{q,z_0,\mu,L\}$ (in the explicit computations we chose $q=L=z_0=1$). Variations of $T/(q\mu)$ can be naturally interpreted as variations of $T$ at constant $\mu$ and $q$. This means that we are specifying $T$ and $\mu$, while the entropy and $\rho$ are dependent variables. In particular notice that if we fix $\mu$ and vary $T$, in general $\rho$ will vary as can be seen by any explicit solution of the equation of motion for $A$. To summarize, we are working in the grand canonical ensemble\footnote{Of course one could insist on using $T/(q\mu)$ \textit{and} to work with the canonical ensemble, i.e.~to keep $\rho$ fixed. In that case variation of $T/(q\mu)$ would be interpreted as some simultaneous variation of $T$ and $\mu$, which has not a very transparent physical interpretation. In the canonical ensemble the results should be naturally plotted as function of $T/q\sqrt{\rho}$. } and we should use the grand canonical potential $\Omega$. 

There are two competing states, the normal state $\psi=0$ and the superconducting state $\psi\neq 0$. We show that below $T_c$ the grand canonical potential of the latter is lower, confirming that a phase transition takes place. We find that the order parameter varies continuously as the temperature crosses $T_c$, while its first derivative does not. Hence the phase transition to the superconducting state is of the second order.
 
By the AdS/CFT dictionary, $\Omega/kT$ is the negative of the on-shell action of the Wick-rotated solution with Euclidean time compactified on a circle of circumference $1/kT$. Very often this quantity is infinite and one needs to renormalize it. We present this derivation for our inhomogeneous holographic superconductor following \cite{Skenderis:2002wp}, but see also \cite{Franco:2009yz,Herzog:2008he,Montull:2009fe} for related computations. We are interested in the action for $A$ and\footnote{For this derivation formulae are simpler if we use $\Psi$ instead of $\psi$. Once we obtain the final result, we will convert it to $\psi=\sqrt{2}\mathrm{Re}\Psi/z$.} $\Psi$ assuming a static and $y$-independent solutions. We will assume that $\Psi$ is real as it is the case for all our solutions. This can be straightforwardly derived from \eqref{La}
\be 
S=\int d^4x\frac12\left[(\partial_z A)^2+\frac{(\partial_x A)^2}{h}\right]-\frac{1}{z^2}\left[h(\partial_z\Psi)^2+(\partial_x\Psi)^2-\Psi^2\left(\frac{A^2 z^2}{h}+2\right)\right]\,.
\ee
\begin{figure}
\centering
\includegraphics[width=5.5in]{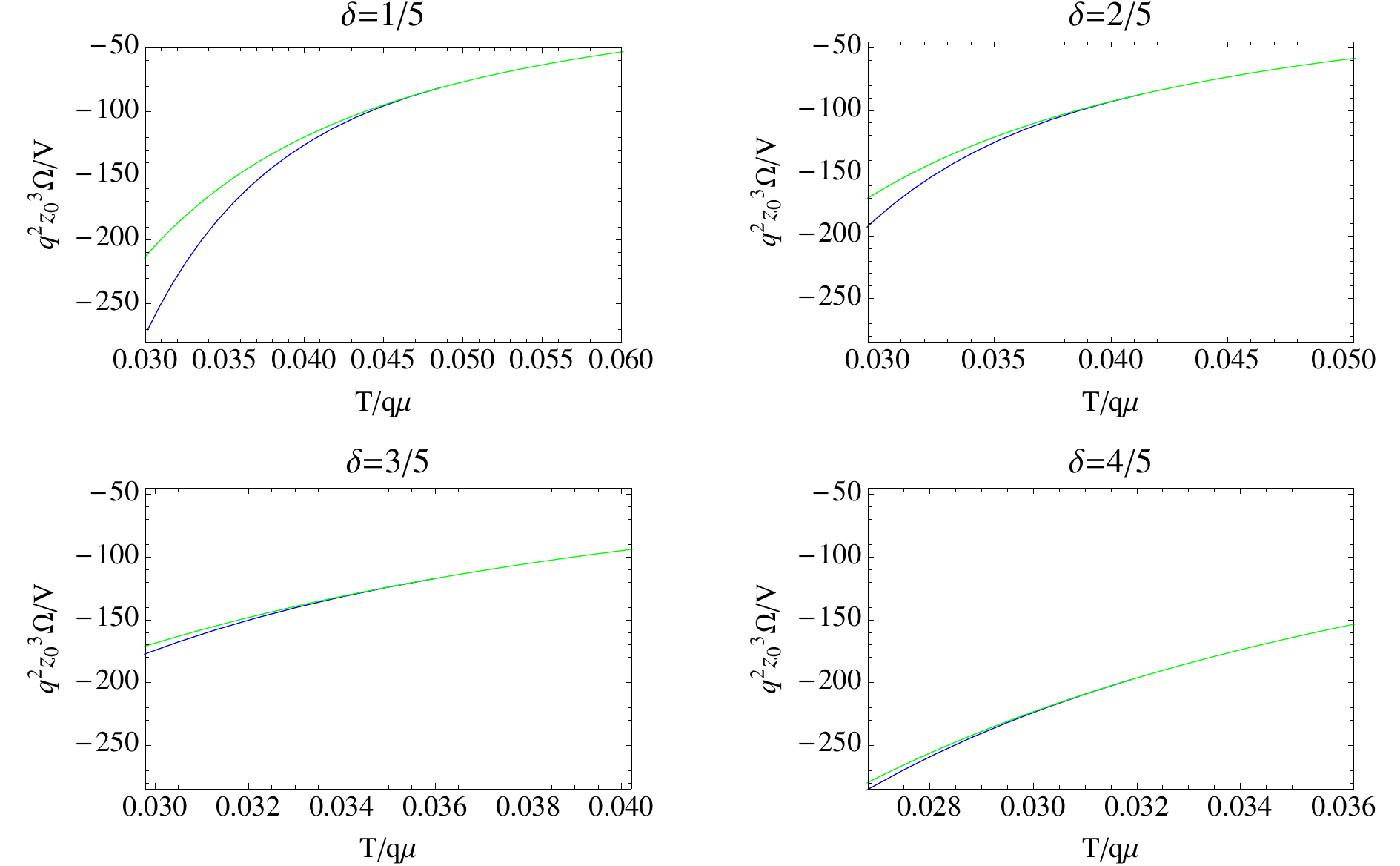}
\caption{The plot shows the grand canonical potential $\Omega$ (multiplied by the appropriate factors to make it invariant under the symmetries in table \ref{t1}) as function of $T/(q\mu)$. Below the critical temperature both the normal and the superconducting states exist. The latter always has a lower $\Omega$.\label{Omegaplot}}
\end{figure}
To evaluate $S$ on shell it is convenient to integrate by parts and use the equations of motion. 
This leads to
\be 
S&=&\int d^4x \left[\mathrm{e.o.m.'s}\right]+\Sos\,,\\
\Sos&\equiv&\int_{z=0} d^3x\left[\frac{h}{z^2}\Psi\Psi'-\frac12 A A'\right]-\int d^4x\frac{A^2}{hz^2}\Psi^2\,,
\ee
where we have used the fact that $A(1,x)=h(1)=0$ while $\Psi(1,x)$ is finite. $\Sos$ has divergences coming from the boundary term at $z=0$. We therefore introduce a cutoff at $z=\epsilon$ to regularize it. The on-shell action can then be expanded as
\be 
\Sos=\int d^3x\frac{[\Psi^{(1)}(x)]^2}{\epsilon}+\mathrm{finite\,terms}\,,
\ee
where we have used the fact that close to the boundary $\Psi$ is given by
\be 
\Psi(z,x)=\Psi^{(1)}(x)z+\Psi^{(2)}(x)z^2+\dots\,.
\ee
This divergence can be subtracted by adding the following boundary counter term
\be 
\Sct\equiv-\int d^3x \frac{h}{z^3}\Psi^2\,.
\ee
Hence the renormalized grand canonical potential is
\be 
\Omega=\int_{z=0} dxdy\left[\frac12 AA'+\frac{h}{z^3}\Psi^2-\frac{h}{z^2}\Psi \Psi'\right]+\int dxdydz\frac{A^2}{h z^2}\Psi^2\,.
\ee
We can simplify this formula when considering only normalizable profiles of $\Psi$, i.e.~$\Psi^{(1)}=0$, which is the case for all the solutions considered in this paper. The average grand canonical potential per unit volume then becomes
\be \label{fen} 
\frac{\Omega}{V}=\frac{Q}{2\pi}\int_{z=0} dx\frac12 AA'+\frac{Q}{2\pi}\int dxdz\frac{A^2}{h z^2}\psi^2\,,
\ee 
where we have reverted to the notation $\psi=\sqrt{2}\mathrm{Re}\Psi/z$. The expression for $\Omega$ further simplifies in the normal state where $\psi=0$. There we can compute \eqref{fen} analytically using our approximate solution for the normal state profile of $A$ 
\be 
 A(z,x)&=&A_0+\au\cos(Qx)\,,\\
 A_0(z)&=&\mu(1-\delta)(1-z)\,,\\
 \au(z)&\simeq&\mu\delta\left[\cosh(Q z)-\coth(Q) \sinh(Q z)\right]\,.
\ee
We find
\be \label{Fan}
\frac{\Omega}{V}=-\frac{1}{2}\mu^2\left[(1-\delta)^2+\frac12\delta^2Q\coth(Q)\right]\,.
\ee
For the superconducting state we have to resort to numerics. It is straightforward, given solutions for $A$ and $\psi$, to numerically evaluate \eqref{fen}. In figure \ref{Omegaplot}, we show the temperature dependence of $\Omega/V$ for several values of $\delta$ for $Q=2$ as an example. In all cases there is a critical value of $T/(q\mu)$ below which two possible solutions exist, the normal one with $\psi=0$ and the superconducting one $\psi\neq0$. For all values of $Q$ and $\delta$ we have studied, the latter has a lower $\Omega$ and is therefore favored. Notice that the numerical computation of $\Omega$ for the normal state is well approximated by the analytical formula \eqref{Fan}.


\section{Conductivity}\label{s:cond}

In this section, we study the electrical conductivity of the boundary theory in the presence of a CDW and both normal and superconducting states. 
Our starting point are the Maxwell equations \eqref{eomA}-\eqref{eomAlast}. We want to add a small homogeneous electric field on top of the solutions we have found in sections \ref{s:Tc} and \ref{s:sc} and study the linear response of the system. Because our solutions are inhomogeneous in the $x$-direction, an electric field in the $x$-direction sources five other independent perturbations even at linear order. The computation of $\sigma_x$ is therefore much more complicated than the computations of conductivity currently performed in the literature. We hope to present results for it in a future publication.

\begin{figure}
\centering
 \includegraphics[width=\textwidth]{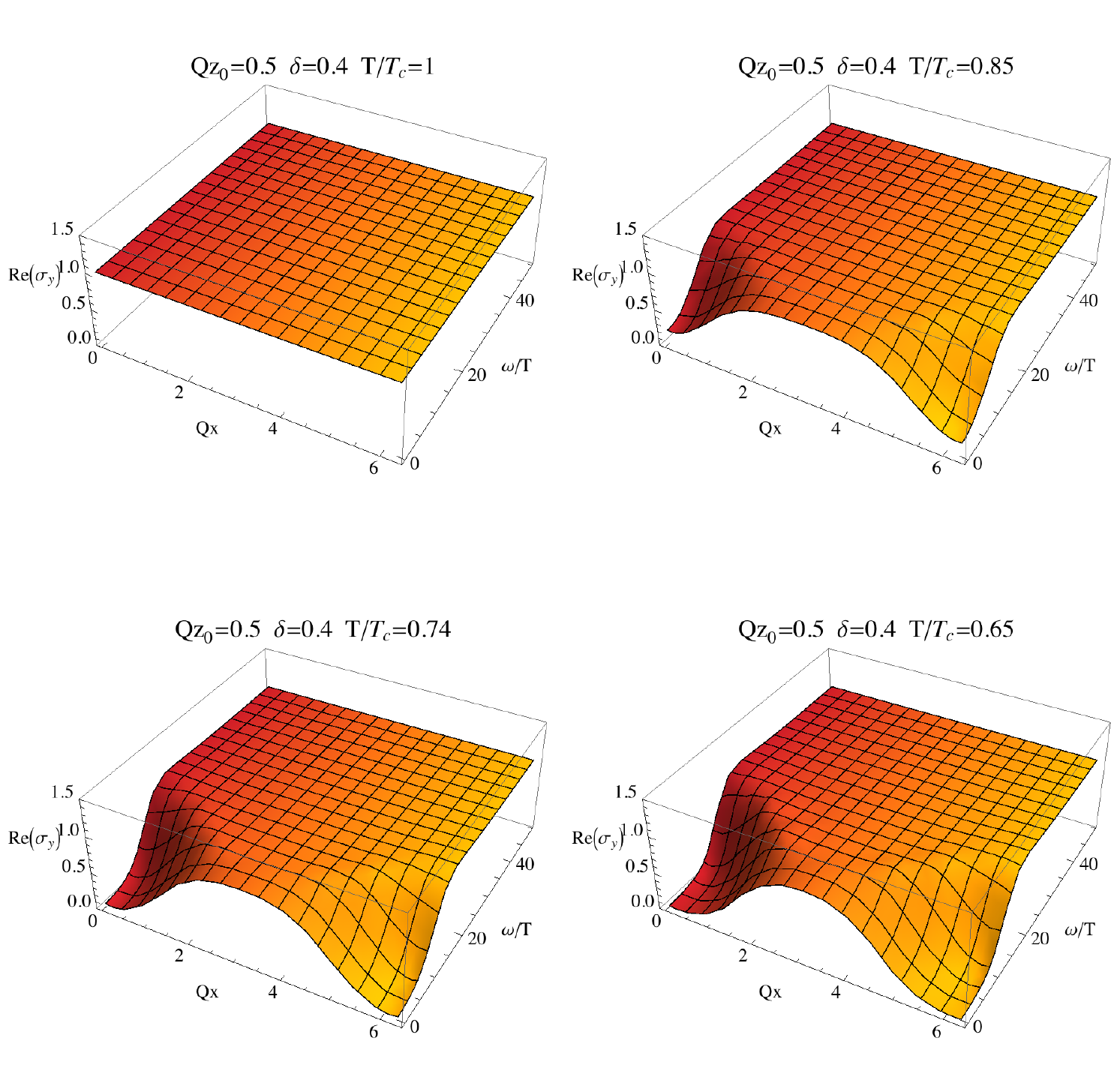}
 \caption{The panel shows the $y$-conductivity $\sigma_y$ as function of $x$ and of the frequency $\omega$ for different values of the temperature and $Q=.5$ and $\delta=.4$. Since $Q$ is small, the system is easily understood in terms of the homogeneous results for the conductivity. A gap opens in correspondence of the superconducting stripes $x=0+2\pi n$, where the order parameter is large (as seen e.g.~in figure \ref{f:stripes}), while $\sigma_y(\omega,x)$ is constant in between the stripes where $\psi\simeq0$. \label{f:cond}}
\end{figure}
\begin{figure}
\centering
 \includegraphics[width=\textwidth]{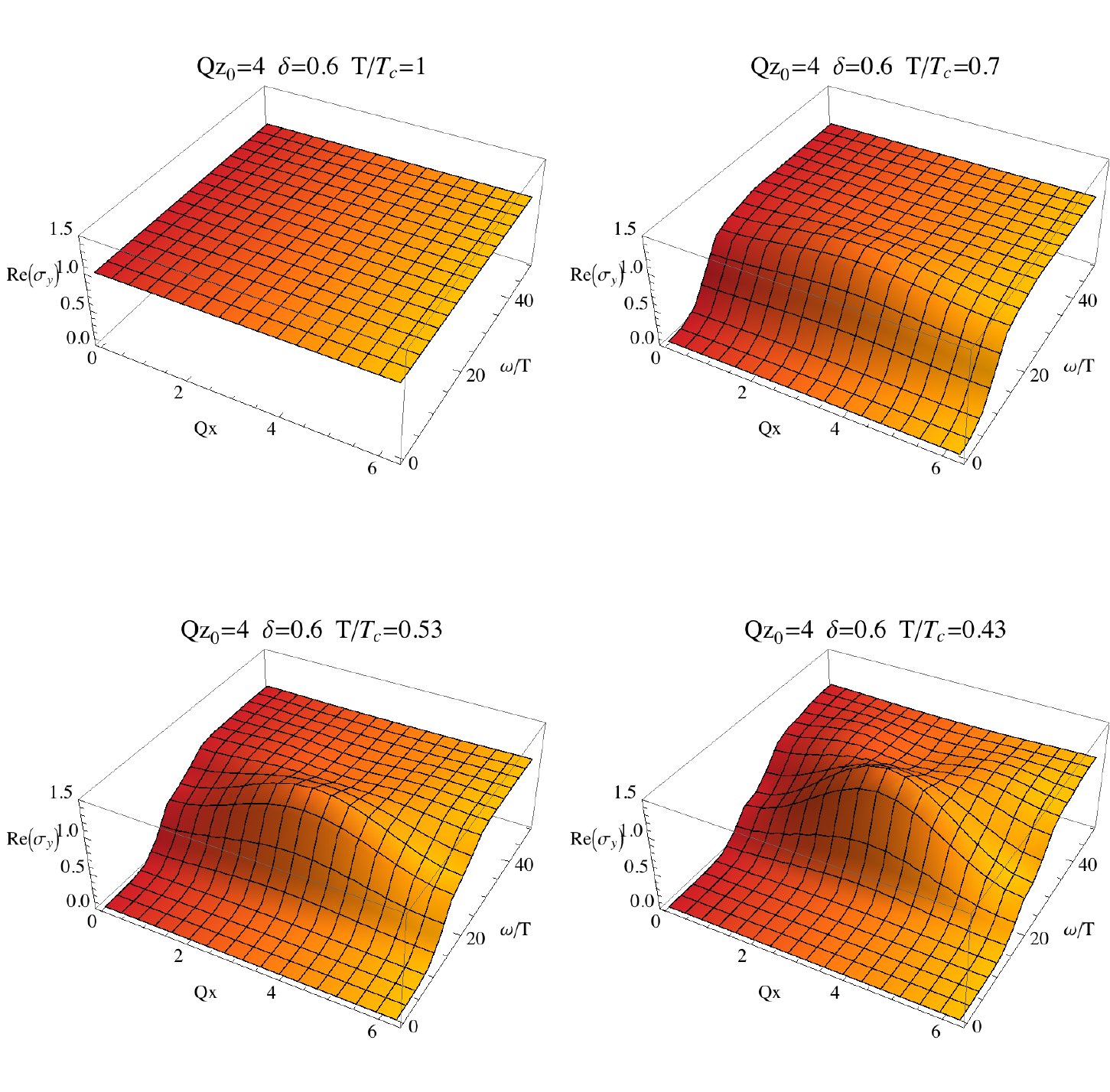}
 \caption{The panel shows the $y$-conductivity $\sigma_y$ as function of $x$ and of the frequency $\omega$ for different values of the temperature and $Q=4$ and $\delta=.4$. The profile is qualitatively very different from the one for $Q<1$ in figure \ref{f:cond}. A gap opens up everywhere in $x$, but the large $\omega$ asymptotic value is reached very differently at different points.\label{f:cond2}}
\end{figure}
\begin{figure}
\centering
 \includegraphics[width=\textwidth]{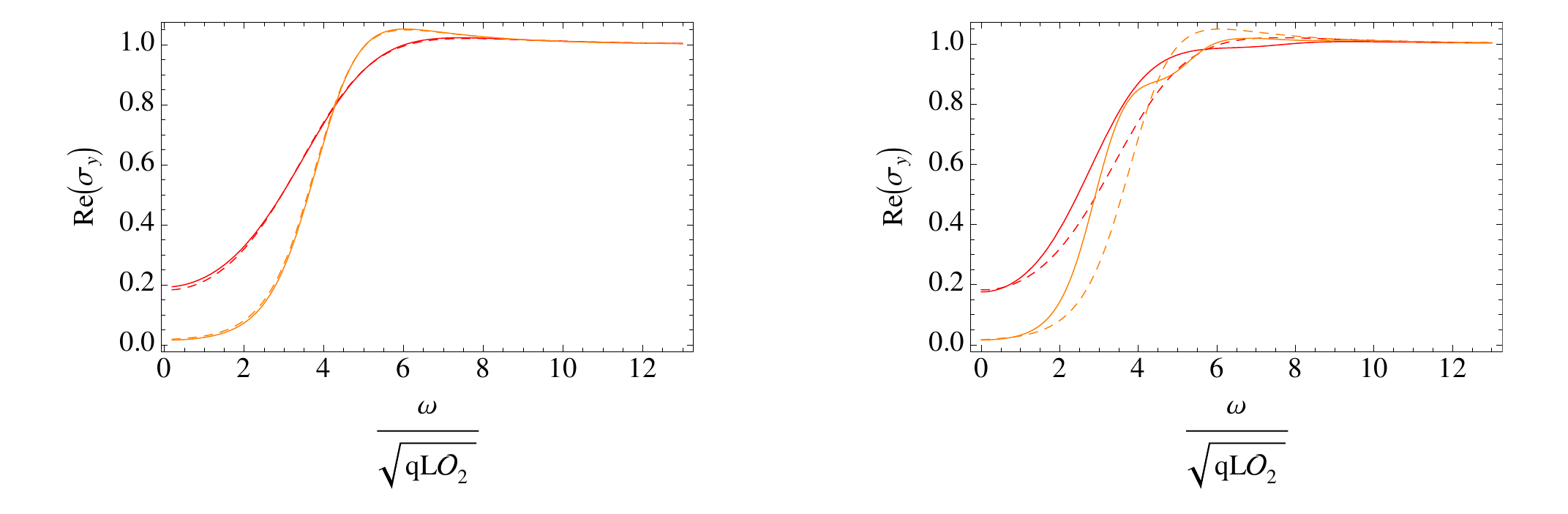}
 \caption{The plots show a comparison of the conductivity for the homogeneous holographic superconductor (dashed line) with an $x=0$ slice of the inhomogeneous conductivity $\sigma_y(\omega,0)$ for two different temperatures $T/T_c=0.85,0.65$. On the left we show $Q=0.5$ while on the right $Q=4$. Once the frequency $\omega$ is rescaled by the value of the corresponding order parameter $\sqrt{\mathcal{O}_2}$, the $Q=0.5$ inhomogeneous conductivity becomes almost indistinguishable from the homogenous one. The $Q=4$ case shows a qualitatively different behavior.\label{f:cond3}}
\end{figure}

The calculation simplifies for $\sigma_y$. In this case, we want to probe the system with a small homogeneous electric field in the $y$-direction, which we will realize by switching on a time dependent $A_y$, and read out the linear response. All the solutions we have presented are homogeneous in the $y$-direction and there are no $\partial_y$ derivatives to contract the perturbation $A_y$ with. Therefore, at linear order in $A_y$, we only have to solve the Maxwell equation for $A_y(\omega,z,x)$ (which is now complex because of the Fourier transform of time into frequency) on the backgrounds computed in the previous sections. Using \eqref{eomAlast} we find
\be \label{cond}
h\partial_z\left(h\partial_z A_y\right)+h\partial_x^2 A_y+\left(\omega^2-\psi^2h\right)A_y=0\,.
\ee
Notice that, since the background is static, the Fourier modes $A_y(\omega,x,z)$ with different frequencies $\omega$ decouple from each other. The optical conductivity in the $y$-direction is given by 
\be \label{def}
\sigma_y(\omega,x)\equiv\frac{J_y(\omega,x)}{E_y(\omega,x)}=-i\frac{A_y^{(1)}(x)}{\omega A_y^{(0)}(x)}\,,
\ee
where we have expanded $A_y(x,z)$ around the AdS boundary $z=0$ as
\be 
A_y(x,z)=A^{(0)}_y(x)+z A_y^{(1)}(x)+\dots\,.
\ee
Let us discuss the two boundary conditions for \eqref{cond}. In the $x$-direction, we impose periodic boundary conditions. In the $z$-direction one boundary condition is simply the overall normalization. As long as $A_y$ is small enough we can work at linear order in $A_y$. The normalization is then irrelevant because it cancels in the definition of $\sigma_y$. The second boundary condition has to be imposed at the horizon $z=z_0=1$ and is dictated by the requirement of causal behavior, which implies ingoing boundary conditions \cite{Son:2002sd}. In our case this is\footnote{In order to see this, plug the ansatz $A_y(\omega,z,x)=(1-z)^{a(\omega)}\tilde A_y(x)+\dots$ into \eqref{cond} and expand around $z=1$. At leading order one finds $a=\pm i \omega/3$. The right sign for ingoing boundary condition, is found by Fourier transforming back to the time domain (this defines our conventions of the sign of the Fourier transform)
\be 
A_y(t,z,x)=\int \frac{d\omega}{2\pi}e^{i\omega \left[t\pm \log(1-z)/3\right]}\tilde A_y(x)+\dots\,.
\ee
Since $\log(1-z)$ decreases going towards the horizon, the ingoing boundary condition is given by the upper sign, i.e.~$A_y(\omega,z,x)\propto(1-z)^{+i\omega/3} +\dots$.} $A_y(\omega,z,x)\propto(1-z)^{+i\omega/3} +\dots$ near the black hole horizon $z=1$.

We have numerically solved \eqref{cond} and computed $\sigma_y$ using \eqref{def}. The results for different values of Q and $T/(q\mu)$ are shown in figures \ref{f:cond} and \ref{f:cond2}. Let us comment on these results starting from the normal CDW background studied in section \ref{s:Tc}. Since $A$ is a vector boson of an Abelian gauge theory, it does not couple to itself and knows about the background only through its coupling to the scalar field. In the normal phase in which the scalar field vanishes, the conductivity in the presence of a CDW is then the same as the one in a homogeneous state, which is $\sigma(\omega)=1$, in the limit in which the backreaction on the geometry is neglected~\cite{Hartnoll:2008kx,Hartnoll:2008vx} (for earlier discussion see \cite{Hartnoll:2007ip}).

Things become more interesting below $T_c$ where a condensate has developed. In figures \ref{f:cond} and \ref{f:cond2}, we show how $\sigma_y(\omega,x)$ changes as the temperature is lowered for two different values of $Q$. Again, two qualitatively different behavior appear depending on the value of $Q$. For $Q<1$, the system effectively behaves as a collection of independent homogeneous holographic superconductors, each one at a fixed value of $x$. This can be formally seen in the analysis of subsection \ref{ss:an} of the $Q\rightarrow 0$ limit. Therefore the conductivity for $Q=.5$ in figure \ref{f:cond} can be qualitatively understood in terms of the conductivity for the homogeneous case. A gap forms for low frequencies in correspondence of the superconducting stripes $x=0+2\pi n$, while in between the stripes, where $\psi\simeq0$, the conductivity is the same as that of the normal state, i.e.~constant. As $Q$ becomes larger than one (keeping all other parameters fixed), there is a smooth crossover to a qualitatively different behavior. We plot the optical conductivity for $Q=4$ in figure \ref{f:cond2}. The profile is dramatically different from the one in figure \ref{f:cond}. A gap opens almost simultaneously everywhere in $x$ due to the increased stiffness in that direction. Interesting resonant patters are visible as the large $\omega$ asymptotic value is reached. To make the new features of the $Q>1$ conductivity even more evident, in figure \ref{f:cond3} we show slices of $\sigma_y(\omega,x)$ at the center of the stripe (i.e.~$x=0$) and compare them with the homogenous conductivity for two subcritical temperatures. Upon rescaling $\omega$ by the corresponding value of the order parameter $\sqrt{\mathcal{O}_2}$, it is clear (left plot of figure \ref{f:cond3}) that the $Q=0.5$ conductivity agrees well with the homogeneous one. On the right plot of figure \ref{f:cond3}, where $Q=4$, $\sigma_y(\omega,0)$ presents extra oscillations which are completely absent in the homogeneous case. It would be desirable to better understand the physics behind the $Q>1$ conductivities.


\section{Conclusions}\label{s:c}

In this paper we have studied a holographic model of superconductivity in the presence of a CDW. Starting from a monochromatic CDW in the normal state with wave vector $Q$, after the phase transition we find superconducting stripes coexisting with the original CDW plus its higher harmonics. The dependence of the temperature of the phase transition $T_c$ on $Q$ has some similarities with what is expected from a weakly coupled BCS computation \cite{martin05}.
$T_c$ decreases monotonically as $Q$ increases, and asymptotes some constant value. The main difference is a steeper functional dependence for $Q\rightarrow \infty$ for the holographic model. These conclusions are derived in the absence of phase fluctuations of the condensate, which is a consistent assumption at the level of our equations. In \cite{martin05} it was argued that once phase fluctuations are included, there is an optimal $Q$ for which $T_c$ has a maximum. It would be very interesting to verify this expectation by generalizing our computation to include these fluctuations. A first step would be a stability analysis of our solutions under small fluctuations. 

To characterize the superconducting phase, we have presented results for the conductivity $\sigma_y$ in the direction parallel to the superconducting stripes. The conductivity in the perpendicular direction, $\sigma_x$ is more complicated to compute because, even at the level of linear response theory, several modes are excited. On the other hand, this is a very interesting observable for a couple of reasons. First, it tells us about the degree of anisotropy of the system, which can be directly compared with experiment. Second, the knowledge of $\sigma_x$ carries information about the correlation between stripes and the nature of proximity effects in a holographic model.

Our analysis has been performed in the probe limit in which the backreaction of the Maxwell and scalar field on the gravitational background are neglected. Although this is expected to capture the relevant physics of the system, it might be interesting to study what happens when gravity is dynamical. One might worry that inhomogeneities might trigger a gravitational collapse making the system unstable. We have two comments in this regard. First, it is well known that stability conditions in AdS are different from and often less intuitive than those in flat space. Second, since $q^2L^2/G_N$ is still a free parameter, one can always make the time-scale of the instability arbitrarily large implying that the system is effectively stable.

There is significant information we could acquire by including the equations for gravity. It would for example be interesting to see whether the DC conductivity perpendicular to the stripes in the normal state has a delta function since translational invariance is broken (as e.g.~in \cite{Hartnoll:2008hs}). One could also see how the findings of \cite{Horowitz:2009ij} about the absence of a hard gap change due to the CDW.

Our focus in this work was to study how superconductivity interacts with a CDW and what the physical consequences of this coexistence are. We did not try to spontaneously generate the inhomogeneity of the CDW solution, but chose to source it with a modulated chemical potential. One could extend our analysis to a model in which the CDW solutions are generated dynamically, e.g. via a spontaneous symmetry breaking along the lines of \cite{Aperis:2010cd}.

We would like to conclude mentioning that we have found several instabilities of the normal state of the holographic superconductor in the presence of a CDW beyond the one studied in this work. Another simple one is towards a superconducting state in which the order parameter is modulated with half the frequency of the CDW. This construction has been proposed in the condensed matter literature to account for various properties of the cuprates and is known as pair density wave (PDW)~\cite{vojta-review}. A thorough discussion of the holographic realization of PDW states will be presented elsewhere.


\section*{Acknowledgments}

We are grateful to Vinay Ambegaokar, Brando Bellazzini, Sean Hartnoll, Jimmy Hutasoit, Ivar Martin, Liam McAllister, James P. Sethna and Jan Zaanen for interesting discussions and correspondence.  The work or R.F. was supported in part by the National Science Foundation under Grant No. NSF-PHY-0747868, the Department of Energy under Grant No. DE-FG02-92ER-40704, and by the World Premier International Research Center Initiative (WPI Initiative), MEXT, Japan. The research of E.P. was supported in part by the National Science Foundation through the grant NSF-PHY-0757868. S.P. acknowledges support by DOE-BES DE-FG02-07ER46393.


\appendix

\section{The Equations of Motion in Components}\label{a:f}

In this appendix, we collect the equations of motion derived from the action \eqref{La}. 

The scalar equation is \cite{Hartnoll:2008kx}
\be\label{eq:eompsi}
- \frac{1}{\sqrt{-g}}D_a\left(\sqrt{-g}g^{ab}D_b\Psi\right) + \frac{1}{2} \frac{\Psi}{|\Psi|} V'(|\Psi|) = 0 \,,
\ee
the equation for the U(1) gauge field is
\be\label{eq:eomA}
\frac{1}{\sqrt{-g}}\partial_a\left(\sqrt{-g} F^{ab}\right) = i qg^{ba} \left[\Psi^* D_a\Psi - \Psi (D_a\Psi)^* \right]
\,,
\ee
and Einstein's equations are
\begin{multline}\label{eq:einst}
R_{ab} - \frac12g_{a b} R - \frac{3}{L^2}g_{ab} =8\pi G_N\left[F_{a c} F_b{}^c - \frac14g_{a b} F^{cd} F_{cd}\right. \\
 \left.- g_{a b}
D_a\Psi(D_b\Psi)^* + \left[D_a\Psi (D_b\Psi)^* + a \leftrightarrow b \right]- g_{a b} V(|\Psi|)\vphantom{\frac14}\right] \,.
\end{multline}
Breaking the complex scalar field up into its real and imaginary part as $\Psi\equiv (\phi+i\chi)/\sqrt{2}$ and using the metric \eqref{metricz}, the equations of motion for the two real scalar fields $\phi$ and $\chi$ become
\be \label{eomphi}
-\frac1{\sqrt{-g}}\partial_a\left(\sqrt{-g}g^{ab}\partial_b\phi\right) + \left(A_aA^a-\frac2{L^2}\right)\phi-\left[\frac1{\sqrt{-g}}\partial_a\left(\sqrt{-g}A^a\right)+2A^a\partial_a\right]\chi &=& 0 \,,\nonumber\\
-\frac1{\sqrt{-g}}\partial_a\left(\sqrt{-g}g^{ab}\partial_b\chi\right) + \left(A_aA^a-\frac2{L^2}\right)\chi+\left[\frac1{\sqrt{-g}}\partial_a\left(\sqrt{-g}A^a\right)+2A^a\partial_a\right]\phi &=& 0 \,.\nonumber
\ee
Analogously, assuming no dependence on the $y$-coordinate, as it is the case for all solutions considered in this paper, the Maxwell equations become
\be \label{eomA}
\frac{z^2}{L^2}\left[h\partial_z^2A_t+\partial_x^2 A_t\right]-A_t(\phi^2+\chi^2)=\hspace{3cm}\nonumber\\
\chi\partial_t \phi-\phi\partial_t \chi+\frac{z^2}{L^2}\left(\partial_t\partial_xA_x+h\partial_t\partial_zA_z\right)\,,\\
\frac{z^2}{L^2}\left[\partial_x^2 A_z-\frac{1}{ h}\partial_t^2A_z\right]-A_z(\phi^2+\chi^2)=\hspace{3cm}\nonumber\\
\chi\partial_z \phi-\phi\partial_z \chi+\frac{z^2}{L^2}\left(\partial_z\partial_xA_x-\frac{1}{h}\partial_t\partial_zA_t\right)\,,\\
\frac{z^2}{L^2}\left[\left(h\partial_z^2 A_x+h'\partial_z A_x\right)-\frac{1}{ h}\partial_t^2A_x\right]-A_x(\phi^2+\chi^2)=\hspace{3cm}\nonumber\\
\chi\partial_x \phi-\phi\partial_x \chi+\frac{z^2}{L^2}\left(h\partial_z\partial_xA_z+h' \partial_xA_z-\frac{1}{h}\partial_t\partial_xA_t\right)\,,\\
\frac{z^2}{L^2}\left[\partial_z\left(h\partial_zA_y\right)-\frac{1}{h}\partial_t^2A_y+\partial_x^2A_y\right]-A_y(\phi^2+\chi^2)=0\,.\label{eomAlast}
\ee
In section \ref{s:Tc} we studied the equation of motion of $\psi$ \eqref{eomphiin} in the normal state background given by
\be
A_{\mathrm{normal}}(z,x)=A_0+ \cos(Qx) \au(z) \equiv \mu (1-\delta)(1-z)+ \cos(Qx) \au(z) \,,
\ee
where $\au$ is the solution of \eqref{simple} with the boundary condition $\au(0)=\delta \mu$ and $\au(1)=0$. Upon expanding $\psi$ in Fourier modes as 
\bea 
\psi(x,z)&=&\sum\limits_{n=0}^\infty \psi_n(z)\cos(n Q x)\,,
\eea
we find an infinite system of linear coupled ordinary differential equations 
\be\label{hie}
&&-\psi_0''+\left[zh(z)-\left(A_0^2+\frac12A_1^2\right)\right]\psi_0-A_0A_1\psi_1-\frac14A_1^2\psi_2=0\,,\nonumber\\
&&-\psi_1''+\left[zh(z)-\left(A_0^2+\frac12A_1^2\right)\right]\psi_1-A_0A_1(2\psi_0+\psi_2)\nonumber\\
&&\hskip 5cm-\frac14A_1^2(\psi_1+\psi_3)+hQ^2\psi_1=0\,,\nonumber\\
&&-\psi_2''+\left[zh(z)-\left(A_0^2+\frac12A_1^2\right)\right]\psi_2-A_0A_1(\psi_1+\psi_3)\nonumber\\
&&\hskip 5cm-\frac14A_1^2(2\psi_0+\psi_4)+h(2Q)^2\psi_2=0\,,\nonumber\\
&&\hskip 6cm\vdots \nonumber\\
&&-\psi_n''+\left[zh(z)-\left(A_0^2+\frac12A_1^2\right)\right]\psi_n-A_0A_1(\psi_{n-1}+\psi_{n+1})\nonumber\\
&&\hskip 5cm-\frac14A_1^2(\psi_{n-2}+\psi_{n+2})+h(nQ)^2\psi_2=0\,,
\ee
where $A_0(z)$ and $A_1(z)$ are the only two Fourier modes of $A$ being non-zero. As an aside, we notice that in the case $A_0=0$ the odd modes $\psi_{2n-1}$ decouple from the even modes $\psi_{2n}$. The full hierarchy for both $A$ and $\psi$ is obtained by plugging the expansions 
\bea 
\psi(x,z)&=&\sum\limits_{n=0}^\infty \psi_n(z)\cos(n Q x)\,,\\
A(x,z)&=&\sum\limits_{n=0}^\infty A_n(z)\cos(n Q x)\,,
\eea
into \eqref{eomphiin} and \eqref{eomAin}. The final form is too long to be written down here, but we give below the Mathematica commands that produce any given order, in this case five
\bea
&&\hskip-1cm{\tt nmax=5;}\\
&&\hskip-1cm{\tt h[z\_]=1-z^3};\nonumber\\[.2cm]
&&\hskip-1cm{\tt{\phi[z\_, x\_] }}={\tt{ Sum[A[n][z] Cos[n Q x], \{n, 0, nmax\}];}}\label{sys2}\nonumber\\
&&\hskip-1cm{\tt{\psi[z\_, x\_] }}= {\tt{Sum[\psi[n][z] Cos[n Q x], \{n, 0, nmax\}];}}\nonumber\\[.2cm]
&&\hskip-1cm{\tt{eqa[Q\_] }}={\tt{ \nonumber
  h[z] D[\phi[z, x], z, z] + 
    D[\phi[z, x], x, x] - \psi[z, x]^2/z^2 \phi[z, x] // 
   TrigReduce; }}\\
&&\hskip-1cm{\tt{eqp[Q\_] }}= {\tt{-h[z] z^2 \nonumber D[h[z]/z^2 D[\psi[z, x], z], z] - 
    2 h[z]/z^2 \psi[z, x] }}\\
    &&\hskip2.5cm{\tt{- 
    h[z] D[\psi[z, x], x, x] - \phi[z, x]^2 \psi[z, x]\nonumber //
   TrigReduce;}}\\[.2cm]
   &&\hskip-1cm{\tt{ Do[\{eqpQ[n][Q\_] }}={\tt{ 
   Collect[If[n == 0, Q/2/Pi Integrate[eqp[Q], \{x, 0, 2 Pi/Q\}], }}\nonumber \\
   &&{\tt{
     \hskip 1cm Coefficient[eqp[Q], Cos[n Q x]]] , \{\psi''[n],\psi'[n],\psi[n]\}, 
    Simplify]\}, \{n, 0, nmax\}] }}\nonumber \\
      &&\hskip-1cm {\tt{ Do[\{eqpQ[n][Q\_] }}={\tt{ 
   Collect[If[n == 0, Q/2/Pi Integrate[eqa[Q], \{x, 0, 2 Pi/Q\}], }}\nonumber \\
   &&{\tt{
     \hskip 1cm Coefficient[eqa[Q], Cos[n Q x]]] , \{A''[n],A'[n],A[n]\}, 
    Simplify]\}, \{n, 0, nmax\}] }}\nonumber
\eea


\end{document}